\documentclass[a4paper,11pt]{article}
\pdfoutput=1 

\usepackage{jcappub} 
\usepackage{soul}
\usepackage[T1]{fontenc} 
\usepackage{leftidx}
\usepackage{xcolor}
\usepackage{lipsum}
\usepackage{mathtools}
\usepackage{cuted}
\usepackage[utf8]{inputenc}
\usepackage{amsmath}
\usepackage{graphicx,bm}

\title{Constraining sterile neutrinos by core-collapse supernovae with multiple detectors}

\author[a]{Jian Tang,}
\author[a,d]{TseChun Wang,}
\author[b,c,d]{Meng-Ru Wu}

\affiliation[a]{School of Physics, Sun Yat-Sen University, Guangzhou 510275, China}
\affiliation[b]{Institute of Physics, Academia Sinica, 
  Taipei, 11529, Taiwan}
\affiliation[c]{Institute of Astronomy and Astrophysics, Academia Sinica, Taipei, 10617, Taiwan}
\affiliation[d]{National Center for Theoretical Sciences, Physics Division, Hsinchu, 30013, Taiwan}

\emailAdd{tangjian5@mail.sysu.edu.cn}
\emailAdd{tsechunwang@mx.nthu.edu.tw}
\emailAdd{mwu@gate.sinica.edu.tw}

\abstract{
The eV-scale sterile neutrino has been proposed to explain some anomalous results in experiments, \textit{such as} the deficit of reactor neutrino fluxes and the excess of $\bar{\nu}_\mu\to\bar{\nu}_e$ in LSND. This hypothesis can be tested by future core-collapse supernova neutrino detection independently since the active-sterile mixing scheme affects the flavor conversion of neutrinos inside the supernova. 
In this work, we compute the predicted supernova neutrino events in future detectors -- DUNE, Hyper-K, and JUNO -- for neutrinos emitted during the neutronization burst phase when the luminosity of $\nu_e$ dominates the other flavors. 
We find that for a supernova occurring within 10 kpc, the difference in the event numbers with and without sterile neutrinos allows to exclude the sterile neutrino hypothesis at more than $99\%$ confidence level robustly.
The derived constraints on sterile neutrinos mixing parameters are comparably better than the results from cosmology and on-going or proposed reactor experiments by more than two orders of magnitude in the $\sin^22\theta_{14}$-$\Delta m_{41}^2$ plane.

}

\begin{document} 
   \maketitle
   \flushbottom
%

\section{Introduction}
The Standard Model (SM) of particle physics is a successful theory in describing properties of known elementary particles. 
However, discovery of neutrino oscillations pointing towards non-zero neutrino masses has been taken as the first direct evidence that SM is not complete.
The framework of three active neutrino mixings has been well-established by solar, atmospheric, accelerator and reactor neutrino oscillation experiments. 
Two mass squared differences ($\Delta m_{21}^2\approx7.4\times10^{-5}$ eV$^2$, and $|\Delta m_{31}^2|\approx2.5\times10^{-3}$ eV$^2$) and three mixing angles ($\theta_{12}\approx 34^\circ$, $\theta_{23}\approx 48^\circ$, $\theta_{13}\approx 8.6^\circ$) are all measured to good precision~\cite{Esteban:2018azc}. 

Furthermore, several reasons indicate that there may exist additional \emph{sterile} neutrinos which have no SM interactions but can mix with active neutrinos or have new interactions via mediators beyond the SM, based on theoretical considerations and experimental hints~\cite{Abazajian:2012ys}. 
Among these, a new sterile neutrino with $\sim$~eV mass mixed with active neutrinos was introduced in late 90's to account for the anomalous results reported by the LSND experiment~\citep{Athanassopoulos:1996jb, Aguilar:2001ty}.
Subsequent experiments like OPERA~\cite{Mauri:2020rua}, KARMEN~\cite{Eitel:1997cy}, MINOS and MINOS$+$ \cite{Adamson:2017uda}, IceCube~\cite{Jones:2019nix}, and MiniBooNE~\cite{Aguilar-Arevalo:2018gpe},
have not yet fully resolved this issue.
In addition to these experiments which measured the appearance of $\nu_e$ from a $\nu_\mu$ (or $\bar\nu_e$ from a $\bar\nu_\mu$) source, measurements of the $\bar\nu_e$ flux near the reactor cores at the baseline of sub-kilometers also reported results in tension with predictions. 
Similarly, light sterile neutrinos at the eV scale are considered as a possible explanation for this tension in the disappearance data of $\bar\nu_e$, dubbed as the ``reactor anomaly''~\cite{Mention:2011rk}. Moreover, the ``gallium anomaly''~\cite{Giunti:2010zu} presented in the $\nu_e$ disappearance data hinted large value of $\sin^22\theta_{14}$, which is currently supported by Neutrino-4 data~\cite{Serebrov:2018vdw} (see, however, Ref.~\cite{Kostensalo:2020hbc}). 
Although the global analysis including both 
the appearance and disappearance data assuming a simple $3+1$ neutrino mixing scheme gives rise to best fits around $|U_{e1}|\sim0.1$ and $|U_{\mu 1}|\sim0.1$ with the new neutrino mass-squared difference $\Delta m_{41}^2\sim 1$~eV$^2$, clear tension among the two data sets remains~\cite{Boser:2019rta, Giunti:2019aiy}.
Further phenomenological models beyond the $3+1$ scheme was reviewed and expected to be assessed by future accelerator as well as reactor neutrino experiments~\cite{Diaz:2019fwt}. 

Complementary to the laboratory searches for sterile neutrinos, their effects on cosmological and astrophysical observation such as the cosmic microwave background (CMB) and the core-collapse supernovae (CCSNe) were also extensively studied in literature. 
In particular, stringent bounds on the mixing parameters between the active and sterile neutrinos based on the inferred effective number of neutrino species $N_{\rm eff}$ from the CMB data were derived, see e.g., Ref.~\cite{Mirizzi:2013gnd,Hagstotz:2020ukm} (see also, however, Ref.~\cite{Hannestad:2013ana,Dasgupta:2013zpn} where secret interactions among sterile neutrinos were included
and Ref.~\cite{Gelmini:2004ah,Hasegawa:2020ctq} which discussed the impact of MeV-scale reheating on the bounds). 
In the context of CCSNe, the mixing of sterile and active neutrinos not only can have a large impact on the explosion mechanism itself and the associated production of heavy elements~\cite{Nunokawa:1997ct,Fetter:2002xx,Tamborra:2011is,Wu:2013gxa,Xiong:2019nvw}, but also on the expected neutrino fluxes of different flavors from a Galactic event~\cite{Esmaili:2014gya,Franarin:2017jnd} as well as the diffuse SN neutrino backgrounds~\cite{Jeong:2018yts}.

Although robust constraints on the mixing parameters based on SNe can not be obtained at present with the detection of only $\simeq 20$ inverse beta-decay (IBD) events from the SN1987a~\cite{Hirata:1987hu,Bionta:1987qt,Alekseev:1988gp}, on-going and upcoming large-size neutrino detectors, including the Deep Underground Neutrino Experiment (DUNE) \cite{Acciarri:2015uup}, Hyper-Kamiokande (Hyper-K) \cite{Hyper-Kamiokande:2016dsw}, and Jiangmen Underground Neutrino Observatory (JUNO)~\cite{An:2015jdp} will record $\mathcal{O}(10^4)$ events when the next Galactic SN occurs. 
More importantly, these detectors can collect SN neutrino events via different interactions. 
For instance, both Hyper-K and JUNO can detect $\bar{\nu}_e$ via the IBD $\bar{\nu}_e+p^+\rightarrow e^+ + n$.
The electron neutrinos $\nu_e$ can be specified by its charged-current (CC) interaction with argon $\nu_e+\leftidx{^{40}}{\text{Ar}}\rightarrow e^-+\leftidx{^{40}}{K^*}$, which is the main detection channel in the liquid-argon detector DUNE. 
The neutral-current (NC) interaction of neutrinos with protons allows JUNO to detect all different flavors of neutrinos and antineutrinos from the proton recoils. 
Finally, the neutrino--electron scattering $e^-+\nu\rightarrow e^-+\nu$ can also produce substantial amount of events in Hyper-K and JUNO.
This motivates us to address in this work how a future Galactic SN can place a quantitative constraint on the mixing between active and eV-scale sterile neutrinos.
In particular, we consider the $3+1$ scenario with a non-zero mixing angle $\theta_{14}$ between the sterile and active neutrinos, relevant for the anomalies in the $\nu_e$ and $\bar\nu_e$ disappearance data.

This paper is organized as follows.
In Sec.~\ref{sec:2}, we introduce the CCSNe and SN neutrinos and discuss the $\nu_e$-$\nu_s$ neutrino flavor conversions inside SNe. 
In Sec.~\ref{sec:simulation}, we describe the detection of SN neutrinos with different interaction channels in DUNE, Hyper-K and JUNO detectors and how we compute the expected events. 
We then use the computed event numbers to derive the constraints from SN neutrino detection on $\sin^22\theta_{14}$ and $\Delta m_{41}^2$ in Sec.~\ref{sec:result}, and discuss various uncertainties.
Conclusion and discussions are given in Sec.~\ref{sec:conclusion}.
 

\section{Supernova neutrinos with $3+1$ mixing}\label{sec:2}

In the final evolutionary stage of a massive star heavier than $\sim 8-10$~$M_\odot$, its innermost core undergoes the gravitational collapse, which then leads to a powerful explosion known as the CCSN and forms a proto-neutron star (PNS).
Neutrinos, being the particles that interact most weakly within the SM, can carry away most of the gravitational binding energy of $\sim 10^{53}$~erg released during such process within $\sim 10$~s,
as qualitatively confirmed by the detection of $\bar\nu_e$ from SN1987a~\cite{Hirata:1987hu,Bionta:1987qt,Alekseev:1988gp}.
This fact together with the CCSN phenomenon has been widely used in the literature to probe some exotic particles beyond the SM such as axions and axion-like-particles~\cite{Raffelt:1987yt,Jaeckel:2017tud,Bar:2019ifz,Carenza:2019pxu}, keV sterile neutrinos~\cite{Raffelt:2011nc,Arguelles:2016uwb,Suliga:2019bsq,Syvolap:2019dat}, dark photons~\cite{Chang:2016ntp,Hardy:2016kme,DeRocco:2019njg,Sung:2019xie}, etc. 
Beyond these, the detection of neutrinos from the next Galactic SN explosion can possibly tell us lots of useful information about the yet-unsolved issues related to CCSNe (see e.g., Ref.~\cite{Horiuchi:2017sku} for a recent review) as well as bring further insights to physics beyond the SM.

Below, we briefly introduce the characteristic emission properties of SN neutrinos and focus on the phase during the neutronization neutrino burst in Sec.~\ref{sec:SNnu}. We then discuss the flavor conversion of SN neutrinos
above the PNS during the neutronization burst phase in Sec.~\ref{sec:nuesconv}.

\subsection{Neutrino emission in core-collapse supernovae}\label{sec:SNnu}

\begin{figure*}
   \centering
   \includegraphics[width=0.48\textwidth]{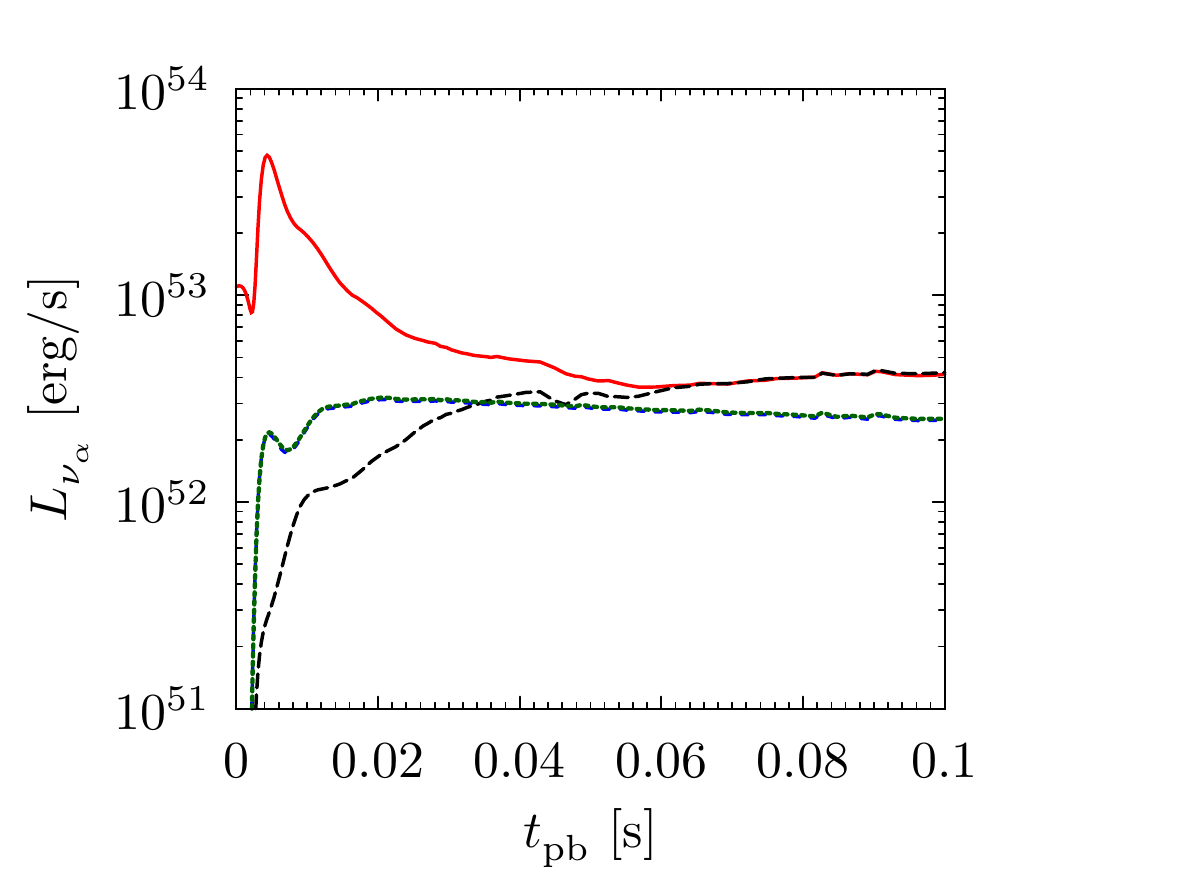}\hspace{-1.2cm}
   \includegraphics[width=0.48\textwidth]{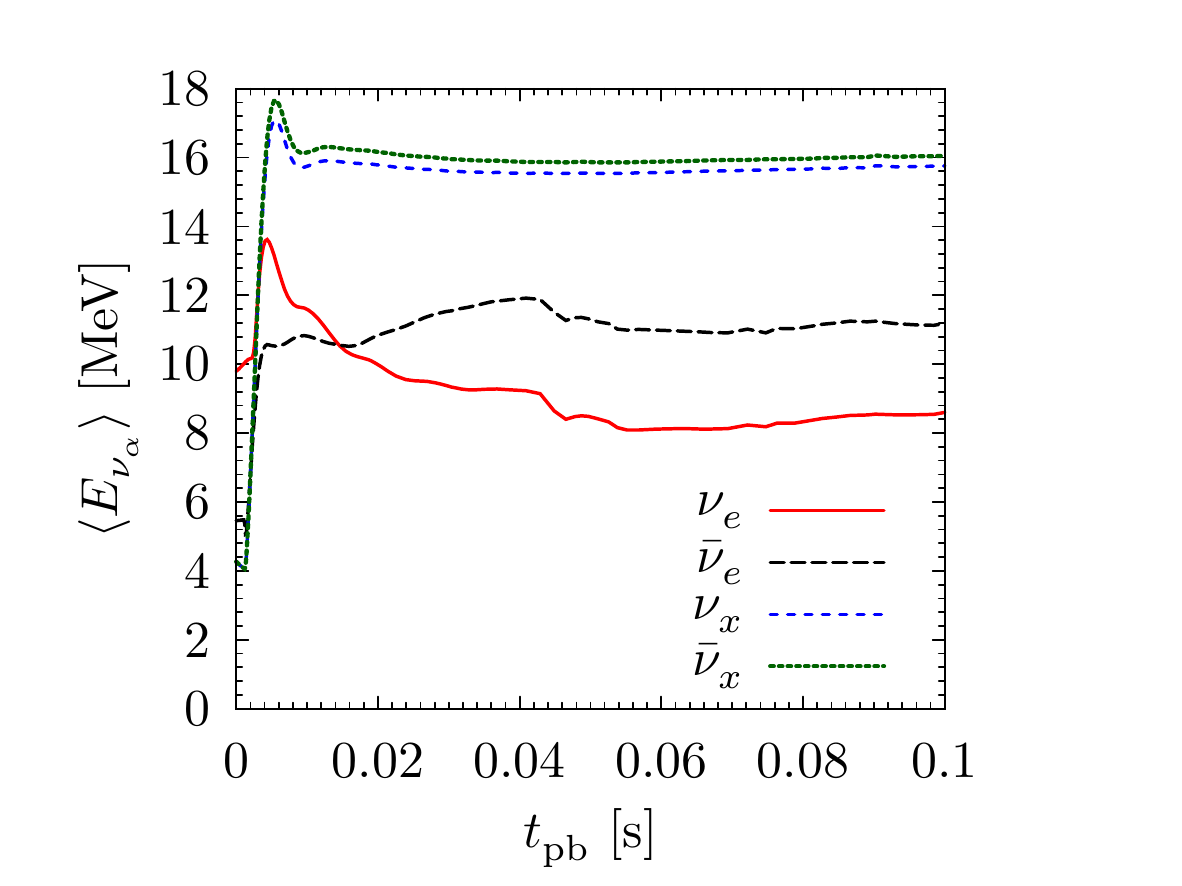}
   \caption{The energy luminosity $L_{\nu_\alpha}$ (left panel) and the mean energy $\langle E_{\nu_\alpha} \rangle$ (right panels) for $\nu_e$, $\nu_x$, $\bar{\nu}_e$, and $\bar\nu_x$ as functions of the post core-bounce time $t_{\rm pb}\leq 0.1$~s adopted in this work, taken from the $18$~$M_\odot$ model in Ref.~\cite{Fischer:2009af}. 
   }
              \label{Fig:flu}%
    \end{figure*}

The emission of neutrinos from CCSNe can be roughly divided into four distinct phases. First, a preSN phase during which neutrinos are emitted via various weak processes before the core bounces. Second, quickly after the core-bounce, a so-called neutronization burst lasting for tens of milliseconds is launched, during which, the $\nu_e$ emission dominates all other flavors (see e.g., Fig.~\ref{Fig:flu}), due to the electron captures on protons dissociated by the SN shock.
Following the neutronization burst, as the SN shock stalls, the continuous mass accretion onto the freshly-formed PNS leads to extended neutrino emission of all flavors for several hundreds milliseconds, with the $\nu_e$ and $\bar\nu_e$ energy luminosity a few times larger than that of other flavors.
Last, after the SN shock being re-launched successfully (likely via neutrino heating), the PNS continues to cool and emits all flavor of neutrinos in roughly equal amounts.
For a recent review, see e.g., Ref.~\cite{Mirizzi:2015eza}.

The mixing of sterile neutrinos with active neutrinos can affect the emergent neutrino flux of all flavors during any of the above phases.
However, we opt to solely focus on the neutronization burst phase throughout this work.
This is because past studies showed that the collective neutrino flavor oscillations, a complicated phenomenon for SN neutrinos (see e.g., Ref.~\cite{Duan:2010bg,Mirizzi:2015eza,Duan:2015cqa} for reviews), can be suppressed during the neutronization burst for a stellar model with an iron core~\cite{Duan:2010bg}, but not at later times.
Thus, we expect that the detection of SN neutrinos during the neutronization burst phase can be used more robustly to probe the $\nu_e$--$\nu_s$ mixing than later phases.

Past studies showed that the normalized SN neutrino energy spectrum $F_{\nu_\alpha}(E_\nu)\equiv dN_{\nu_\alpha}/dE_\nu$ for a flavor $\alpha$ at a given time can generally be well-fitted by a 
pinched Fermi-Dirac distribution characterized by the mean energy $\langle E_{\nu_\alpha}\rangle$ and the pinching parameter $\gamma_{\nu_\alpha}$~\cite{Keil:2002in,Tamborra:2012ac},

\begin{equation}\label{eq:fluence_2}
F_{\nu_\alpha}(E_\nu)\equiv\frac{dN_{\nu_\alpha}}{dE_\nu}(E_\nu)=A_{\nu_\alpha}\left(\frac{E_\nu}{\langle E_{\nu_\alpha}\rangle}\right)^{\gamma_{\nu_\alpha}}\exp\left[-\left(\gamma_{\nu_\alpha}+1\right)\frac{E_\nu}{\langle E_{\nu_\alpha}\rangle}\right],
\end{equation}
where $A_{\nu_\alpha}=(\gamma_{\nu_\alpha}+1)^{\gamma_{\nu_\alpha}+1}/(\langle E_{\nu_\alpha}\rangle\Gamma(\gamma_{\nu_\alpha}+1))$ is the normalization constant with $\Gamma(\gamma_{\nu_\alpha}+1)$ being the Gamma function, and 
\begin{equation}
\gamma_{\nu_\alpha}=\frac{2\langle E_{\nu_\alpha} \rangle^2-\langle E_{\nu_\alpha}^2 \rangle}{\langle E_{\nu_\alpha}^2 \rangle -\langle E_{\nu_\alpha} \rangle ^2}.
\end{equation}
Fittings to numerical simulations often result in $\gamma_\nu\simeq 3$~~\cite{Keil:2002in,Tamborra:2012ac}.

Fig.~\ref{Fig:flu} shows the time evolution of the 
energy luminosity $L_{\nu_\alpha}$ for different neutrino flavors and their $\langle E_{\nu_\alpha} \rangle$ computed in Ref.~\cite{Fischer:2009af} with an $18~M_\odot$ progenitor\footnote{We use $\nu_x$ to denote both $\nu_
\mu$ and $\nu_\tau$ as they have almost the same energy spectrum when emitted from the PNS.}. 
It shows clearly that during the first 20~ms post the core-bounce, the energy luminosity of $\nu_e$ is much larger than that of the other flavors. 
The $L_{\nu_e}$ peaks at $\simeq 4$~ms and has a width of $\simeq 10~$ms. The mean energy of all flavors also increase rapidly during the first $\sim 5$~ms before settling into a clear hierarchy of 
$\langle E_{\nu_e}\rangle < \langle E_{\bar\nu_e}\rangle < \langle E_{\nu_x}\rangle\simeq \langle E_{\bar\nu_x}\rangle$.

Since we are interested in computing the total neutrino events at Earth detectors integrating over the neutronization burst phase, we define the neutrino fluence at Earth as

\begin{equation}\label{eq:fluence_1}
\Phi_{\nu_\alpha}(E_\nu)=\frac{1}{4\pi D^2}\frac{\epsilon_{\nu_\alpha}}{\widehat{\langle E_{\nu_\alpha}\rangle}}\widehat{\frac{dN_{\nu_\alpha}}{dE_\nu}}(E_\nu),
\end{equation}
where $D$ is the distance between the SN and the Earth and $\epsilon_{\nu_\alpha}=\int_{t_0}^{t_1} dt L_{\nu_\alpha}(t)$ is the total energy carried by $\nu_\alpha$ from time $t_0$ to $t_1$.
The time-averaged mean energy $\widehat{\langle E_{\nu_\alpha} \rangle}$ can be computed as
\begin{equation}\label{eq:mean_energy_over_time}
\widehat{\langle E_{\nu_\alpha} \rangle}=   \frac{\int^{t_1}_{t_0} dt L_{\nu_\alpha}(t)}{\int^{t_1}_{t_0} dt L_{\nu_\alpha}(t)/\langle E_{\nu_\alpha} \rangle(t)},
\end{equation}
and $\widehat{\frac{dN_{\nu_\alpha}}{dE_\nu}}(E_\nu)$
has the same form as in Eq.~\eqref{eq:fluence_2} by replacing both the ${\langle E_{\nu_\alpha} \rangle}$ and $\gamma_{\nu_\alpha}$ by $\widehat{\langle E_{\nu_\alpha} \rangle}$ and $\hat\gamma_{\nu_\alpha}=\int_{t_0}^{t_1} dt \gamma(t)/(t_1-t_0)$, correspondingly.

\begin{table}[b!]
\centering
\begin{tabular}{c|c|c|c}
\hline\hline
 & $\epsilon_{\nu_\alpha}$~[$10^{50}$~erg] & $\widehat{\langle E_{\nu_\alpha}\rangle}$~[MeV] & $\hat{\gamma}_{\nu_\alpha}$ \\\hline
 $\nu_e$ & $24.8$ & $12.3$ & $3$ \\\hline
$\bar{\nu}_e$ & $0.396$ & $10.6$ & $3$ \\\hline
$\nu_x$ & $1.36$ & $15.7$ & $3$ \\\hline
$\bar{\nu}_x$ & $1.39$ & $16.1$ & $3$ \\\hline\hline
\end{tabular} 
\caption{\label{tab:fluence_para}The default fluence parameters used in this work (see Eq.~\eqref{eq:fluence_1} and text below for definitions).}
\end{table}

Table~\ref{tab:fluence_para} lists the computed values of $\epsilon_{\nu_\alpha}$, $\widehat{\langle E_{\nu_\alpha} \rangle}$, and $\hat\gamma_{\nu_\alpha}$ using the luminosities and mean energies shown in Fig.~\ref{Fig:flu} during the first 10~ms post-bounce ($t_0=0$ and $t_1=10$~ms). 
For simplicity, we assume time-independent $\gamma_{\nu_\alpha}=3$.
We use these values as our default model for most analyses and discuss effects due to the uncertainties of the neutrino fluence in Sec.~\ref{sec:fluuncer}\footnote{We note already here that we have tested another choice of $t_1=20$~ms. However, it leads to nearly identical results in terms of the exclusion limits discussed in Sec.~\ref{sec:eventnumber}.}

\subsection{$\nu_e$-$\nu_s$ conversion in supernovae}\label{sec:nuesconv}

The flavor evolution of a neutrino is governed by a Schr\"odinger-like equation with an effective Hamiltonian which consists of a vacuum term $H_{\rm v}$ due to the flavor mixing and a matter potential matrix $V$ due to the forward scattering of neutrinos with particles in medium \cite{Mikheev:1986gs,Mikheev:1986wj,Wolfenstein:1979ni},
\begin{align}
i\frac{d}{dt}\Psi_\nu=(H_{\rm v}+V)\Psi_\nu,
\label{eq:hamiltonian}
\end{align}
where the $\alpha$ component of $\Psi_\nu$, $\Psi_{\nu,\alpha}$
denotes the amplitude of a neutrino in an eigenstate $\nu_\alpha$.
Considering the $3+1$ sterile-active mixing scenario, 
the vacuum Hamiltonian matrix in flavor basis can be written as 
\begin{equation}
H_{\rm v}=
\frac{U M^2 U^\dagger}{2E_\nu}=    
\frac{1}{2E_\nu}\left(
\begin{array}{cccc}
m_{ee}^2 & m_{e\mu}^2 & m_{e\tau}^2 & m_{es}^2 \\
m_{\mu e}^2 & m_{\mu\mu}^2 & m_{\mu\tau}^2 & m_{\mu s}^2 \\
m_{\tau e}^2 & m_{\tau\mu}^2 & m_{\tau\tau}^2 & m_{\tau s}^2 \\
m_{s e}^2 & m_{s\mu}^2 & m_{s\tau}^2 & m_{s s}^2 
\end{array}
\right),
\end{equation}
where $U$ is the vacuum mixing matrix 
and the mass-squared matrix $M^2=\text{diag}(0,\Delta m_{21}^2,\Delta m_{31}^2,\Delta m_{41}^2)$.
Neglecting the neutrino-neutrino forward scattering term as they only contribute subdominantly during the neutronization burst phase of concern, the potential matrix $V$ for neutrinos is given by
\begin{equation}
V=\left(
\begin{array}{cccc}
V_{\text{CC}}-V_{\text{NC}} & 0 & 0 & 0\\
0& -V_{\text{NC}} & 0 & 0\\ 
0& 0& -V_{\text{NC}} & 0\\
0& 0& 0& 0
\end{array}
\right),    
\label{eq:potential_1}
\end{equation}
where $V_{\text{CC}}\equiv \sqrt{2}G_\text{F}N_e E_\nu$ and $V_{\text{NC}}\equiv \sqrt{2}G_\text{F}N_n E_\nu/2$ with $N_e$ ($N_n$) being the net electron (neutron) number density, and $G_{\rm F}$ the Fermi constant.
Note that for antineutrinos, the corresponding potential $\bar V$ differs from $V$ by a minus sign.
Due to charge neutrality, $N_n$ can be related to $N_e$ via the electron number fraction $Y_e\equiv N_e/N_b$ ($N_b$ is the baryon number density) by 
\begin{equation}
\frac{N_n}{N_e}=\frac{1}{Y_e}-1.
\end{equation}
Thus, Eq.~\eqref{eq:potential_1} becomes, 
\begin{equation}
\begin{array}{l}
V=V_{\text{CC}}\left(
\begin{array}{cccc}
\frac{3Y_e-1}{2Y_e} & 0 & 0 & 0\\
0& \frac{Y_e-1}{2Y_e} & 0 & 0\\ 
0& 0& \frac{Y_e-1}{2Y_e} & 0\\
0& 0& 0& 0
\end{array}
\right).
\end{array}
\end{equation}
The Hamiltonian in Eq.~\eqref{eq:hamiltonian} determines the mixing between the mass and flavor eigenstates in medium, as the mixing matrix diagonalizes the Hamiltonian.
It is obvious that when the matter-effect potential $V$ dominates the Hamiltonian for $2E_\nu\times V_{\rm CC}\gg \Delta m^2_{41}$, the in-medium mass eigenstates and the mixings are defined only by the potential matrix $V$, i.e. the flavor and in-medium mass states nearly coincide. 

When neutrinos traverse outward from the PNS surface where $V$ dominates, their in-medium mass eigenstates change as the matter density $\rho$ and $Y_e$ vary in the stellar environment. 
As we focus on the effect of sterile neutrinos in flavor conversions assuming only a non-zero $\theta_{14}$ ($\theta_{24}=\theta_{34}=0$),
one can define the Mikheyev-Smirnov-Wolfenstein (MSW) resonance condition in the 1--4 subspace when
\begin{equation}\label{eq:res}
\cos2\theta_{14}=\frac{2V_{ee}E_\nu}{\Delta m^2_{41}}=\frac{2\sqrt{2}E_\nu G_\text{F}N_e(3Y_e-1)}{2Y_e\Delta m_{41}^2},
\end{equation}
where we have defined $V_{ee}=V_{\rm CC}(3Y_e-1)/(2Y_e)$.
The neutrino flavor conversion across the resonance is adiabatic if the change of the potential $V_{ee}$ is slow enough to satisfy the condition,
\begin{equation}\label{eq:gamma}
\gamma_{14}\equiv\frac{(\Delta m_{41}^2)^2\sin^22\theta_{14}}{4E_\nu^2}{\left|\frac{dV_{ee}}{dr}\right|^{-1}_{\text{res}}}\gg 1,
\end{equation}
where the subscript $_{\text{res}}$ denotes that the value is being evaluated at the resonance defined in Eq.~\eqref{eq:res}. 
The adiabatic parameter $\gamma_{14}$ defines the transition probability of a neutrino from an in-medium mass eigenstate $\nu_1$ to $\nu_4$ (or vise versa) when going through the resonance by 
\begin{equation}
p_{14}\simeq\exp(-\pi\gamma_{14}/2).
\end{equation}
Clearly, an adiabatic transition ($\gamma_{14}\gg 1$) implies $p_{14}\simeq 0$.

   \begin{figure*}
   \centering
   \includegraphics[width=0.48\textwidth]{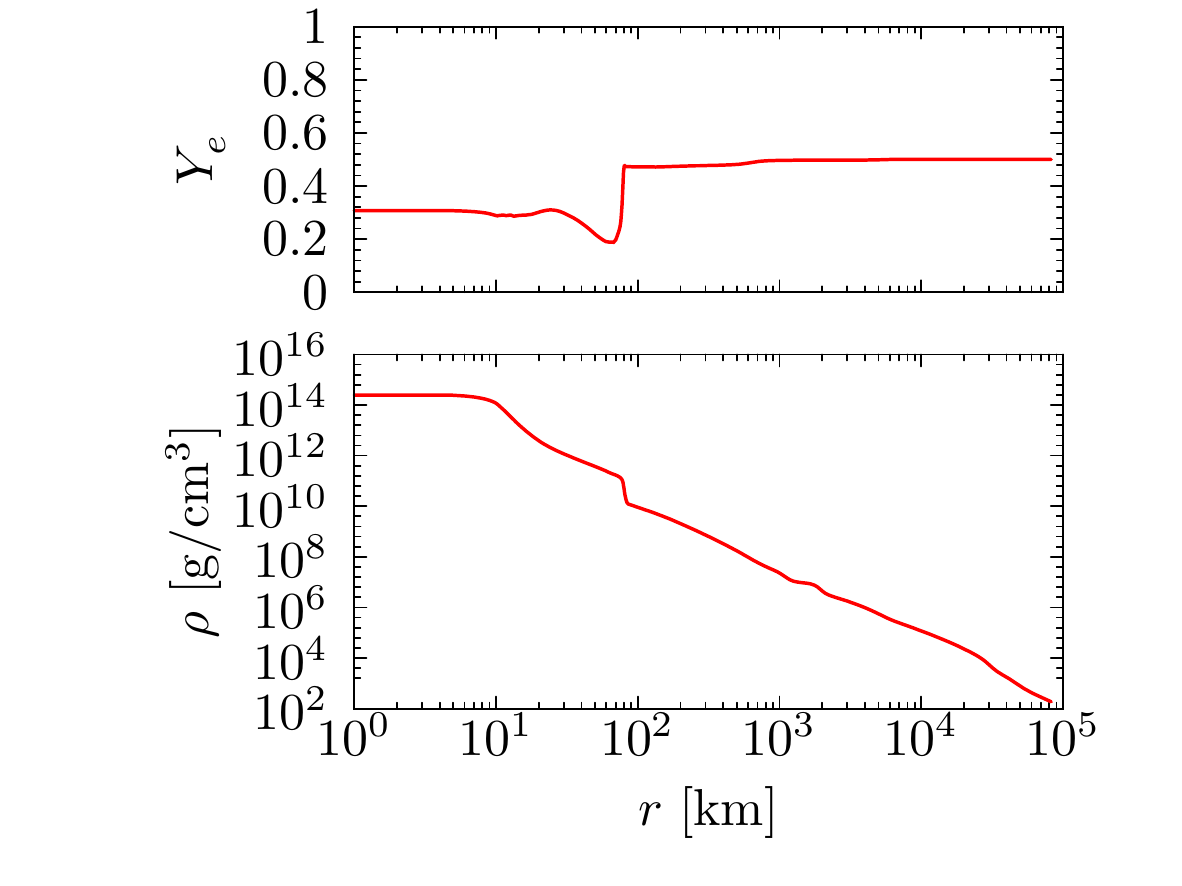}
   \includegraphics[width=0.48\textwidth]{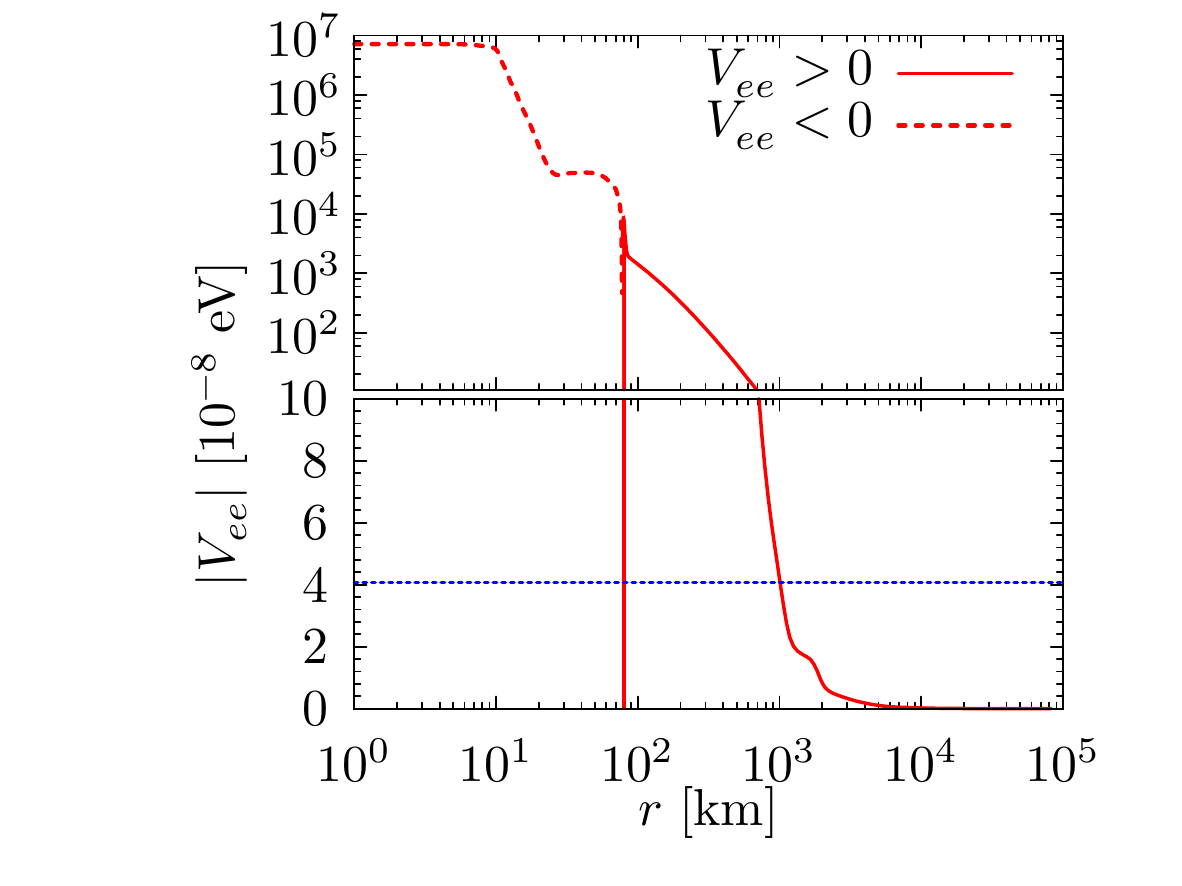}
   \caption{Left: The matter density $\rho$ (lower) and electron number fraction per baryon $Y_e$ (upper) as functions of radius $r$ in a SN core, during the neutronization burst phase.
   The profiles are taken from the 18~$M_\odot$ model in Ref.~\cite{Fischer:2009af} at a post-bounce time of $4.37$~ms. Right: The absolute value of matter effect potential $|V_{ee}|$ against $r$. The solid and dashed red curves represent positive and negative values of $V_{ee}$, respectively.
   The horizontal dotted blue line indicates the value of $\Delta m^2_{41}/(2E_\nu)$ for $\Delta m_{41}^2=1$~eV$^2$ and $E_\nu=\widehat{\langle E_{\nu_e}\rangle}=12.3$~MeV.
   }
              \label{Fig:profile}%
    \end{figure*}

In Fig.~\ref{Fig:profile}, we show in the left panel the density $\rho(r)$ and $Y_e(r)$ profiles at $t_{\rm pb}=4.37$~ms, the peak-time of the neutronization burst (see Fig.~\ref{Fig:flu}) and the corresponding $V_{ee}$ in the right panel. 
We have checked that during the neutronization phase $\rho$ and $Y_e$ profiles do not change significantly.
The quantity $\Delta m_{41}^2/(2E_\nu)$ for $\Delta m_{41}^2=1$~eV$^2$ and $E_\nu=\widehat{\langle E_{\nu_e}\rangle}=12.3$~MeV are also shown together with $V_{ee}$.
One can see that for SN neutrinos, the resonance conditions in the 1--4 subspace are satisfied at two locations: The inner resonances at where $Y_e\simeq 1/3$ for both neutrinos and antineutrinos, as well as an outer resonance for neutrinos further out at where $\rho\sim 10^6$~g$\cdot$cm$^{-3}$.
During the neutronization burst, the inner resonance locates at where the SN shock is. The shock discontinuity guarantees that flavor conversions for $\nu_e\rightarrow\nu_s$ and $\bar\nu_e\rightarrow\bar\nu_s$ when passing through their inner resonances are extremely non-adiabatic, i.e., $\gamma_{14}\rightarrow 0$, and can be ignored hereafter. 
Thus, one can characterize the flavor conversion probability of a $\nu_e$ to $\nu_s$ by a single $p_{14}$ computed at the outer resonance.
Hereafter, we use exclusively $p_{14}$ to denote the transition probability of neutrinos at the outer resonance.
We note that as shown in the right panel of Fig.~\ref{Fig:profile}, for a $\Delta m^2_{41}$ larger than few hundred eV$^2$, the resonance condition for $\nu_e$ will no longer be satisfied. Thus, we restrict our analysis to $\Delta m^2_{41}\leq 100$~eV$^2$.

After the outer resonance, the flavor evolution of all neutrino species simply follow the well-known adiabatic flavor conversions~\cite{Dighe:1999bi}. 
Thus, one can explicitly express the normalized neutrino spectra 
taking into account the effect of flavor conversions, $F_{\nu_\alpha}$, in terms of the values without oscillations, $F^0_{\nu_\alpha}$.
For the normal ordering (NO) of the neutrino masses ($m_3>m_2>m_1$),
including the $\nu_e$--$\nu_s$ mixing gives
\begin{subequations}\label{eq:Fe_normal}
\begin{align}
F_{\nu_e}= &
\left\{|U_{e3}|^2p_{14}+|U_{e4}|^2(1-p_{14})\right\}F_{\nu_e}^0
+\left(|U_{e1}|^2+|U_{e2}|^2\right)F_{\nu_x}^0, \\
F_{\nu_x}= &
(|U_{\mu 3}|^2+|U_{\tau 3}|^2)p_{14}F_{\nu_e}^0  +\left(|U_{\mu 1}|^2+|U_{\tau 1}|^2+|U_{\mu 2}|^2+|U_{\tau 2}|^2\right)F_{\nu_x}^0, \\
F_{\bar{\nu}_e}= & |U_{e1}|^2F^0_{\bar{\nu}_e}+\left(|U_{e2}|^2+|U_{e3}|^2\right)F^0_{\bar{\nu}_x}, \\
F_{\bar{\nu}_x}= & (|U_{\mu 1}|^2+|U_{\tau 1}|^2)F^0_{\bar{\nu}_e}+\left(|U_{\mu 2}|^2+|U_{\tau 2}|^2+|U_{\mu3}|^2+|U_{\tau3}|^2\right)F^0_{\bar{\nu}_x},
\end{align}
\end{subequations}
where $U_{\alpha i}$ are the $\alpha i$'s component in the mixing matrix $U$ ($\alpha$ is the flavor index for $e$, $\mu$, $\tau$ and $s$, and $i=1\cdots4$).
Note that we have set $|U_{\mu 4}|=|U_{\tau 4}|=0$ and the flavor conversion of $\bar\nu_e\rightarrow\bar\nu_s$ at the inner resonance can be ignored as discussed above.

It is straightforward to verify that Eqs.~\eqref{eq:Fe_normal} reduce to the formulas derived by assuming only adiabatic flavor conversions with three active-neutrino mixings~\cite{Dighe:1999bi}, when taking $|U_{e4}|=0$ and $p_{14}=1$:
\begin{subequations}\label{eq:Fe_normal_3nu}
\begin{align}
F_{\nu_e}= & |U_{e3}|^2 F^0_{\nu_e}+ (1-|U_{e3}|^2) F^0_{\nu_x},\\
F_{\nu_x}= & (1-|U_{e3}|^2) F^0_{\nu_e}+ (1+|U_{e3}|^2) F^0_{\nu_x}, \\
F_{\bar{\nu}_e}= & |U_{e1}|^2F^0_{\bar{\nu}_e}+(1-|U_{e1}|^2)F_{\bar{\nu}_x}^0,\\
F_{\bar{\nu}_e}= & (1-|U_{e1}|^2)
F^0_{\bar{\nu}_e}+(1+|U_{e1}|^2)F_{\bar{\nu}_x}^0.
\end{align}
\end{subequations}

Comparing Eqs.~\eqref{eq:Fe_normal} with Eqs.~\eqref{eq:Fe_normal_3nu}, it indicates the following important consequence.
For $F^0_{\nu_e}\gg F^0_{\nu_x}$, the $\nu_e$ flux at Earth, $F_{\nu_e}$, is always suppressed regardless of whether the $\nu_e\rightarrow\nu_s$ flavor conversion is adiabatic or not for $\theta_{14}\ll 1$ due to the small values of $|U_{e3}|^2\approx0.02$. 
However, for the $\nu_x$ flux at Earth, $F_{\nu_x}\simeq 0$ when $p_{14}\rightarrow 0$,
while $F_{\nu_x}\simeq F^0_{\nu_e}$ without sterile neutrinos.
Thus, it is crucial to include detectors capable of detecting SN $\nu_x$ to probe the $\nu_e$--$\nu_s$ mixing.
For the case of inverted mass ordering (IO) of active neutrinos, the corresponding equations are given in Appendix~\ref{sec:flacon_IO}.
They imply once again that considering detectors capable of detecting SN $\nu_x$ is important, as in the case of NO.

\begin{figure*}[t!]
   \centering
   \includegraphics[width=0.48\textwidth]{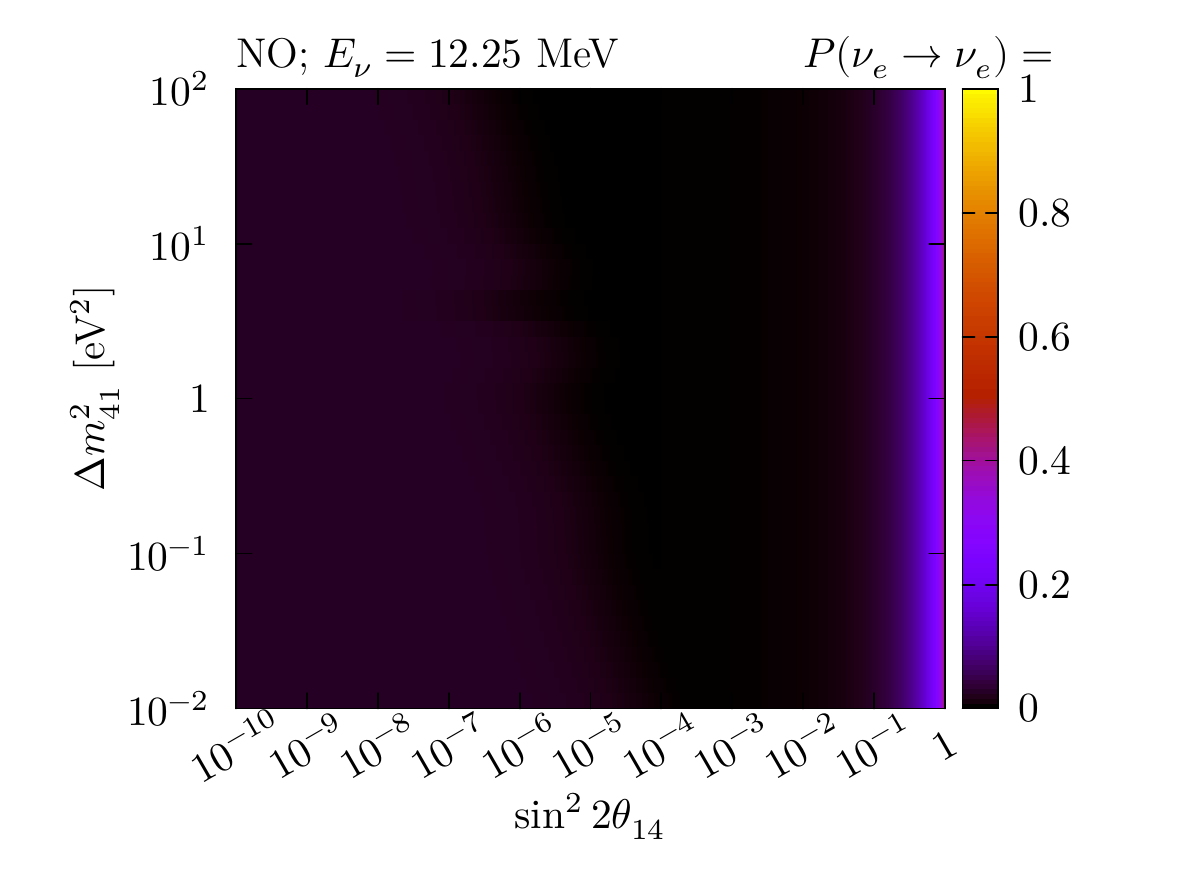}\hspace{-0.5cm}
   \includegraphics[width=0.48\textwidth]{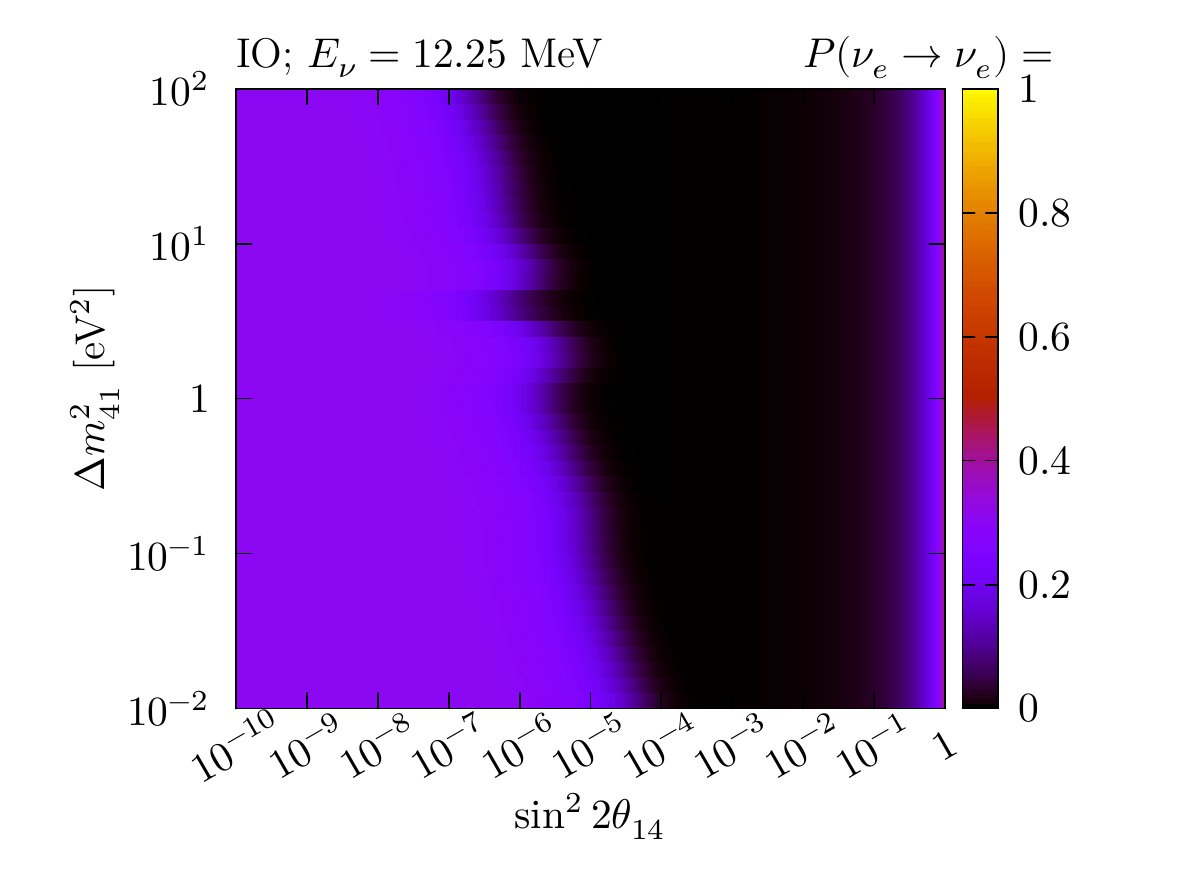}
   
      \includegraphics[width=0.48\textwidth]{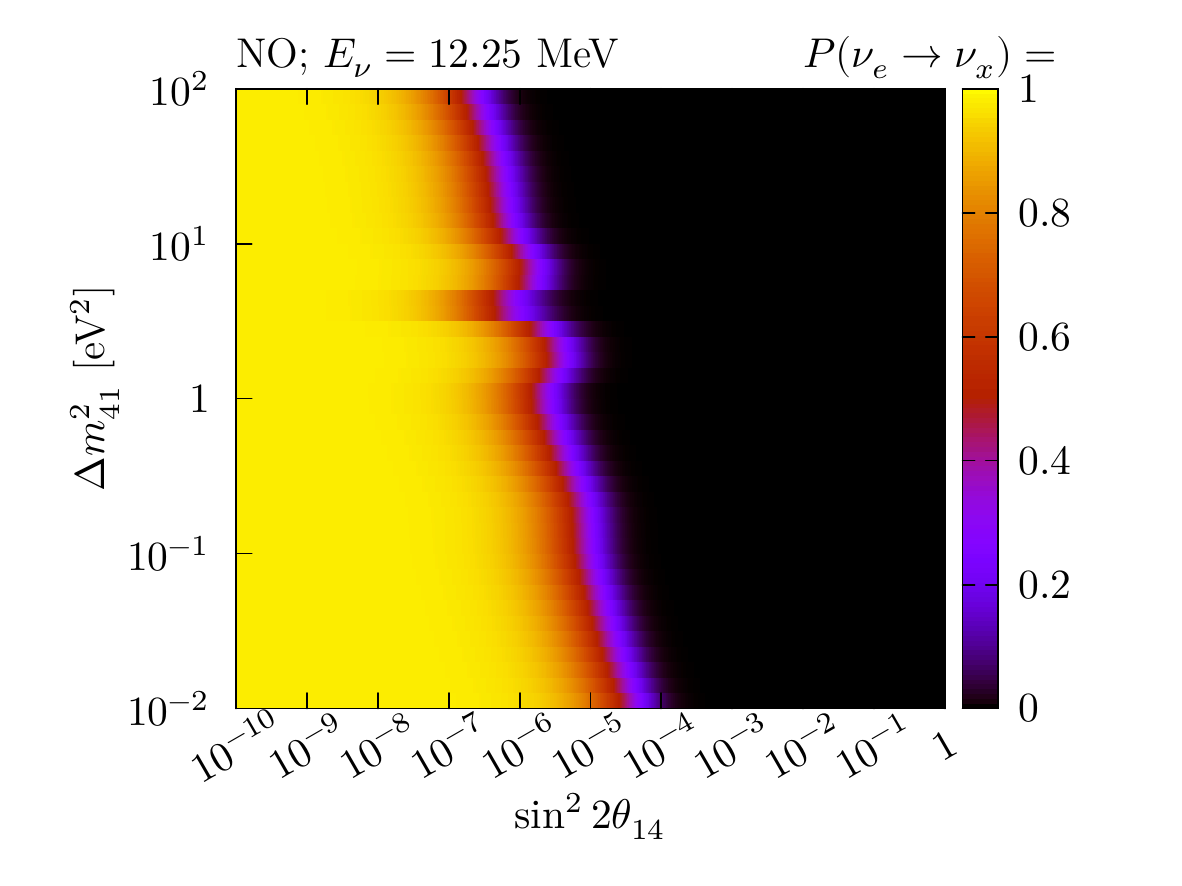}\hspace{-0.5cm}
   \includegraphics[width=0.48\textwidth]{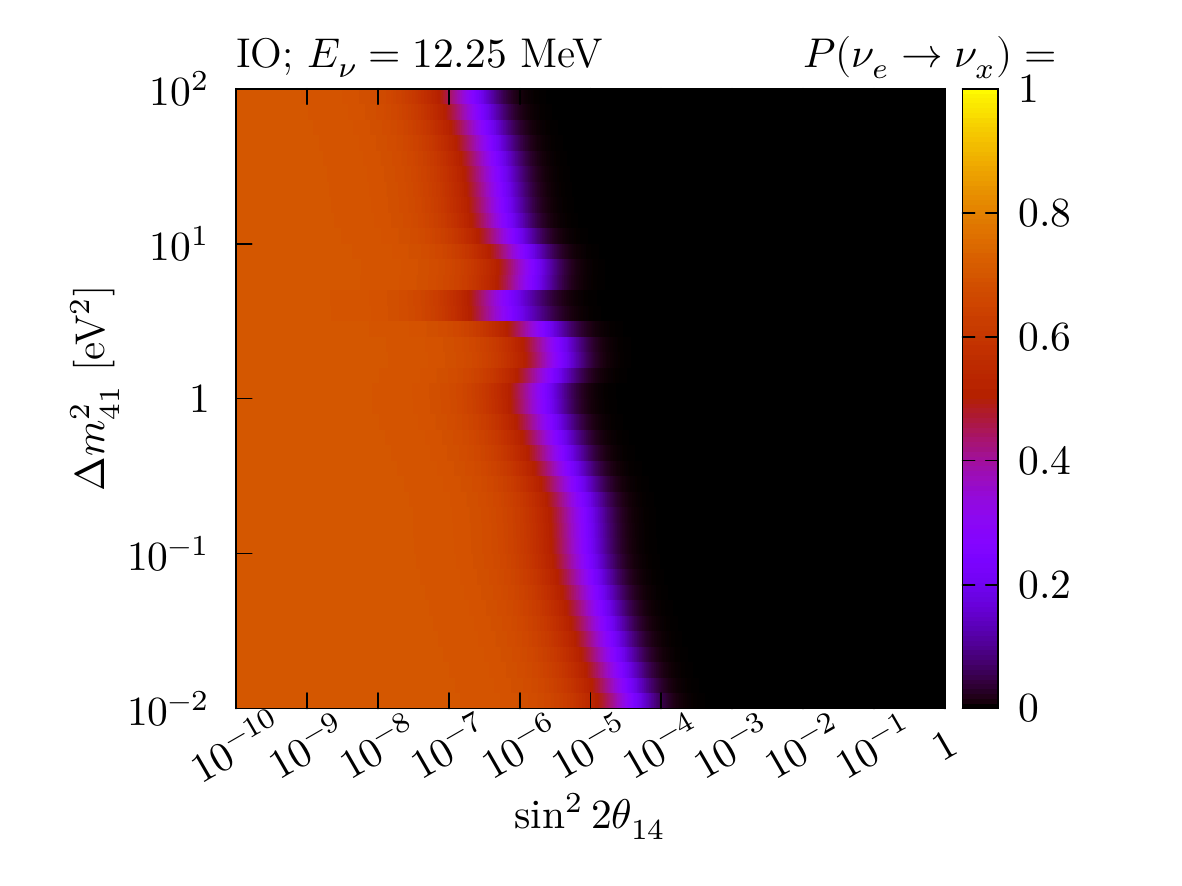}
   \caption{The flavor survival probability $P(\nu_e\rightarrow\nu_e)$ (upper panels) and conversion probability $P(\nu_e\rightarrow\nu_x)$ (lower panels) for a $\nu_e$ with energy $E_\nu=\widehat{\langle E_{\nu_e}\rangle}=12.3$~MeV traversing through the entire SN, in the $\nu_e$--$\nu_s$ mixing parameter space ranging from $10^{-2}~\text{eV}^2<\Delta m_{41}^2<10^2~\text{eV}^2$ and $10^{-10}<\sin^22\theta_{14}<1$, for $\theta_{24}=\theta_{34}=0$.
   Left and right panels are for the NO and IO cases, respectively.
   The matter density and the electron fraction used for this figure is shown in Fig.~\ref{Fig:profile}. 
   }
   \label{Fig:prob}
\end{figure*}

In Fig.~\ref{Fig:prob}, we show the flavor conversion probabilities of $\nu_e\rightarrow\nu_e$,
\begin{equation}
P(\nu_e\rightarrow\nu_e)=|U_{e3}|^2p_{14}+|U_{e4}|^2(1-p_{14}),
\end{equation}
and that of $\nu_e\rightarrow\nu_x$,
\begin{equation}
P(\nu_e\rightarrow\nu_x)=(|U_{\mu 3}|^2+|U_{\tau 3}|^2)p_{14},
\end{equation}
as functions of $\Delta m^2_{41}$ and $\sin^2{2\theta_{14}}$ taking $E_\nu=\widehat{\langle E_{\nu_e} \rangle}=12.3$~MeV for both NO and IO, computed with $\rho(r)$ and $Y_e(r)$ profiles shown in Fig.~\ref{Fig:profile}.
They clearly show that over the wide range of $10^{-2}~\rm{eV}^2\leq \Delta m^2_{41} \leq 10^2$~eV$^2$, the $\nu_e\rightarrow\nu_s$ flavor conversion is almost adiabatic for $\sin^2{2\theta_{14}}\gtrsim 10^{-5}\times(\rm{eV}^2/\Delta m^2_{41})^{1/2}$ (see Eq.~\eqref{eq:gamma})\footnote{We note that the small wiggle around $\Delta m^2_{41}\simeq$ a few eV$^2$ is due to the density bump located at $r\simeq 10^3$~km (see Fig.~\ref{Fig:profile}).}.
This results in a $P(\nu_e\rightarrow\nu_e)=|U_{e4}|^2\simeq 0.25\sin^2\theta_{14}$ for $\theta_{14}\ll 1$, and $P(\nu_e\rightarrow\nu_x)= (|U_{\mu 3}|^2+|U_{\tau 3}|^2)p_{14} \simeq 0$. Both values are different from those predicted assuming the $3$--$\nu$ mixing paradigm where $P(\nu_e\rightarrow\nu_e)\simeq 0.02~(0.3)$ and 
$P(\nu_e\rightarrow\nu_x)\simeq 0.98~(0.7)$ for NO (IO).
Note that for a large $\sin^2{2\theta_{14}}\gtrsim 0.1$, 
$P(\nu_e\rightarrow\nu_e)$ rapidly increases with $\sin^2{2\theta_{14}}$ and approach 0.5 for $\sin^2{2\theta_{14}}=1$.
For smaller $\sin^2{2\theta_{14}}\lesssim 10^{-5}\times(\rm{eV}^2/\Delta m^2_{41})^{1/2}$, the flavor conversion of $\nu_e\rightarrow\nu_s$ quickly becomes non-adiabatic such that the resulting $P(\nu_e\rightarrow\nu_e)$ and $P(\nu_e\rightarrow\nu_x)$ approach the values given by the $3$--$\nu$ mixing paradigm. Based on these results, we expect that a future galactic SN can possibly constrain the $\nu_e$--$\nu_s$ mixing down to very small mixing angle of $\sin^2{2\theta_{14}}\sim 10^{-5}\times(\rm{eV}^2/\Delta m^2_{41})^{1/2}$ given enough event statistics, for which we address in the following sections.


\section{Supernova neutrino detection} \label{sec:simulation}
Neutrinos can be detected by means of weak interactions. Both charged-current (CC) and neutral-current (NC) interactions can contribute to the SN neutrino detections. It is essential to turn the recoiled energy into electromagnetic signals by photomultipliers. Elastic scatterings on electrons, reserve the directional information if a Cherenkov cone with regards to angular and recoil energy distributions of the scattered electrons can be well reconstructed as was established by water Cherenkov detectors~\cite{Fukuda:1998mi, Beacom:1998ya}. The channel of elastic scatterings between electrons and its partner neutrino $\nu_e$ is more promising while the channel $\bar{\nu}_e-e^-$ is helicity suppressed. The muon- and tau-flavor (anti)neutrino signals from a SN can only be extracted on the statistical basis after we subtract the electron (anti)neutrino contributions from the measured NC interactions since their energy is too low to create partner muons or taus in CC interactions. The alternative promising technique is the measurement of the recoil energy from scatterings between neutrinos and protons via NC interactions~\cite{Beacom:2002hs}. The proton gets the maximal kinetic energy when the neutrino direction is completely reverted after such a collision. Then the incoming neutrino energy can be reconstructed based on the recoiled proton kinetic energy. Liquid scintillator detectors such as JUNO have lots of free protons in the hydrogen atom and can help with SN neutrino detections. Of course, neutrinos also take part in interactions with nucleons in nuclei by CC and NC interactions but the cross sections are somewhat smaller than those in interactions with free protons. Quenching effects have to be included to correct the nuclear effects. Besides, neutrino interactions at $\mathcal{O}(10)$ MeV on nuclei are poorly understood in theory. The typical uncertainties in theory reach the level of $10\%-20\%$. The angular and recoil energy distributions of the subsequent daughters are lost more often than not to facilitate reconstructions of neutrino signals. Needless to say, these interactions have very different thresholds and energy-dependent cross sections due to deexcitations on different target nuclei. For example, the CC neutrino-nucleus interactions on $^{12}$C, $^{13}$C and $^{13}$N for galactic SN neutrinos were considered in future large liquid scintillator detectors~\cite{Lu:2016ipr}. In addition, the MeV-scale neutrinos can take part in coherent neutrino-nucleus scatterings with relatively higher cross sections but suffer from the extremely low threshold due to such a small recoil energy. Therefore, we focus on neutrino and free proton interactions in the liquid scintillator detector, and leave the bound nucleon interactions out of the current study. Nevertheless, in liquid argon time projection chamber~\cite{Amerio:2004ze} we take a different strategy by means of  $\nu_e+\leftidx{^{40}}{\text{Ar}}\rightarrow e^-+\leftidx{^{40}}{K^*}$, where deexcitation signals from $\leftidx{^{40}}{K^*}$ can be tagged in a high efficiency.

In summary, we expect to collect SN neutrino signals by the proton elastic scatterings $\nu+p^+\rightarrow \nu+p^+$ ($p$ES) as expected at JUNO, electron elastic scatterings $\nu+e^-\rightarrow \nu+e^-$ ($e$ES) to be reconstructed in JUNO and Hyper-K, and the CC interactions with Argon $\nu_e+\leftidx{^{40}}{\text{Ar}}\rightarrow e^-+\leftidx{^{40}}{K^*}$ interactions (ArCC events) by DUNE. Electron antineutrino events from IBD provide prompt light via positron-electron annihilation and delayed light via the deexcitations from the captured neutron. IBD is a good detection channel and provides a large amount of $\bar{\nu}_e$ during the whole CCSN explosion, yet electron antineutrinos are relatively fewer and show much less significance than electron neutrinos during the neutronization burst, as shown in Fig.~\ref{Fig:flu}. For this reason, we put IBDs aside for JUNO and Hyper-K. Due to the smaller cross sections for neutrino-nucleus interactions on $^{12}$C, $^{13}$C and $^{13}$N compared with $e$ES and $p$ES, we also suppose that their contributions are negligible here.
The detection processes and their most sensitive neutrino flavors in each detector are summarised in Table~\ref{tab:exp_event}.

\begin{table}[b!]
\centering
\caption{\label{tab:exp_event}The considered detection channels in this work.}
\begin{tabular}{c|c|c|c}
\hline\hline
Experiment & Channels & Detector & Sensitive neutrino flavors\\\hline
DUNE &   ArCC & $40$-kton Liquid Argon & $\nu_e$\\
Hyper-K & $e$ES  &$370$-kton Water & $\nu_e,~\nu_x,~\bar{\nu}_e,~\bar{\nu}_x$\\
JUNO  & $e$ES and $p$ES& $20$-kton Liquid Scintillator& $\nu_e,~\nu_x,~\bar{\nu}_e,~\bar{\nu}_x$\\
\hline\hline
\end{tabular} 
\end{table}

The potential backgrounds for SN neutrino signals could come from close-by reactor power plants ($\bar{\nu}_e$), solar neutrinos ($\nu_e$), low-energy atmospheric neutrinos ($\nu_e$, $\bar{\nu}_e$, $\nu_\mu$ and $\bar{\nu}_\mu$), cosmic muons, $\alpha$ and $\beta$ decays from radioactive isotopes in target media and the surrounding materials in each detector. We expect instant and high event rates from SN neutrinos relative to reactor, solar and atmospheric neutrinos. Overburden for a neutrino detector in the underground laboratory will suppress the cosmic muon induced backgrounds. Radiopurity control of detector components will make the contributions from radioactive isotopes negligible. We also note that inefficient neutron captures in the IBD prompt events might pollute $p$ES and $e$ES but remain as sub-leading effects comprised by detector-related systematic uncertainties.
For simplicity, we assume that the backgrounds can be neglected. 

\subsection{$e$ES events}
The Cherenkov detector takes the ultrapure water as the target material. The $e$ES channel 
can be registered. Charged particles can be identified with the reconstructed Cherekov cones. The directional information can be well reconstructed but we have to bear with the relatively high threshold and low energy resolution due to low light yield from the Cherenkov process compared with the liquid scintillator technology. The total event rate $N_{\text{eES}}$ based on $e$ES is given by
\begin{equation}\label{eq:eES_total}
N_{\text{eES}}=\int^{E_\text{max}}_{E_\text{th}} \frac{dN_{\nu e}}{dE_o} d E_o,
\end{equation}
where $E_o$ is the visible energy observed by photomultipliers. We set the threshold E$_\text{th}=0.2$ MeV and the cut-off energy E$_\text{max}=60$ MeV. The differential event rate $\frac{dN_{\nu e}}{dE_o}$ can be expressed as follows: 
\begin{equation}\label{eq:N_eES}
\frac{dN_{\nu e}}{dE_o}=N_e\sum_\alpha\int^\infty_0 dT_e\cdot G(E_o;T_e,\delta_E)
\times\int^\infty_{E_{\alpha}^{\text{min}}}\frac{dF_{\alpha}}{dE_\alpha}\cdot\frac{d\sigma_{\nu e}(E_\alpha)}{dT_e}dE_\alpha,
\end{equation}
where $N_e$ is the total number of electrons in the detector, $T_e$ is the recoil energy for electrons, and $E^{\text{min}}_{\alpha}$ is the minimal neutrino energy $\approx T_e/2+\sqrt{T_e(T_e+2m_e)}/2$. The function $G(E_o;T_e,\delta_E)$ is a Guassian distribution for the energy resolution with the central value $T_e$ and the standard width $\delta_E$. In our simulation, we only consider the total event number, and therefore this factor $G(E_o,T_e,\delta_E)$ is less relevant in our discussion. We set its width at $3\%$ for simplicity.

We use the cross section for eES interactions,
\begin{equation}
\frac{d\sigma_{\nu e}(E_\nu)}{dT_e}=\frac{2m_e G_\text{F}^2}{\pi}\left[\epsilon_-^2+\epsilon_+^2\left(1-\frac{T_e}{E_\nu}\right)^2-\epsilon_-\epsilon_+\frac{m_eT_e}{E_\nu^2} \right].
\end{equation}
Here $G_\text{F}=1.166\times 10^{5}$ GeV$^{-2}$, and the expressions of $\epsilon_+=-\sin^2\theta_{W}$, and $\epsilon_-=-1/2-\sin^2\theta_{W}$ for $\nu_e$ or $\epsilon_-=1/2-\sin^2\theta_{W}$ for $\nu_x$. For the antineutrino mode, we just have to swap $\epsilon_-$ and $\epsilon_+$.

\subsection{$p$ES events}
The typical liquid scintillator detector takes the hydrocarbons as the target material. Recoil energy is transferred to scintillation light to be registered by photo-multipliers with the advantages of good energy resolution and low thresholds. However, the direction information for the incoming SN neutrinos can hardly be reconstructed well since the scintillation light is isotropic. 
To get the total $p$ES event number $N_{p\text{ES}}$, we do the integration
\begin{equation}\label{eq:N_pES}
N_{p\text{ES}}=\int^{E_\text{max}}_{E_\text{th}} \left[\int^{\infty}_{0}\frac{dN_{\nu p}}{dT'_{p}}\times G(E_o,T'_p,\delta_E) dT'_p\right] d E_o,
\end{equation}
where $E_o$ is the visible energy observed by photo-sensors and $T'_p$ is the recoil energy of the final-state proton after quenching. We set the threshold E$_\text{th}=0.2$ MeV and the cut-off energy E$_\text{max}=60$ MeV. The $p$ES event rate in $T_p'$ can be evaluated by 
\begin{equation}\label{eq:pES_dNdT}
\frac{dN_{\nu p}}{dT'_{p}}=N_p\sum_\alpha\frac{dT_p}{dT'_p}\int^{\infty}_{E^{min}_\alpha} \frac{dF_\alpha}{dE_\alpha}\frac{d\sigma_{\nu p}(E_\alpha)}{dT_p}dE_\alpha,
\end{equation}
where $T_p$ is the recoil energy of the final-state proton before quenching. We require the neutrino energy above $E^{min}_\alpha=\sqrt{T_pm_p/2}$. The number of $p$ES events is not small due to the contributions of both $\nu_\mu$ and $\nu_\tau$ and their antiparticles, whose average energy is higher than those of $\nu_e$ and $\bar{\nu}_e$. $N_p$ is the number of protons in the detector, according to \cite{Li:2017dbg}, this value is approximately $1.4\times10^{33}$ in JUNO.
The differential energy-dependent spectrum in Eq.~\eqref{eq:pES_dNdT} will be further convoluted with the energy resolution function $G(E_o, T^\prime_p, \delta_E)$ to give the event spectrum with respect to the observed energy $E_o$, as we discussed before in Eq.~\eqref{eq:N_eES}.

\begin{figure}[h!]
\centering 
\includegraphics[width=0.6\textwidth]{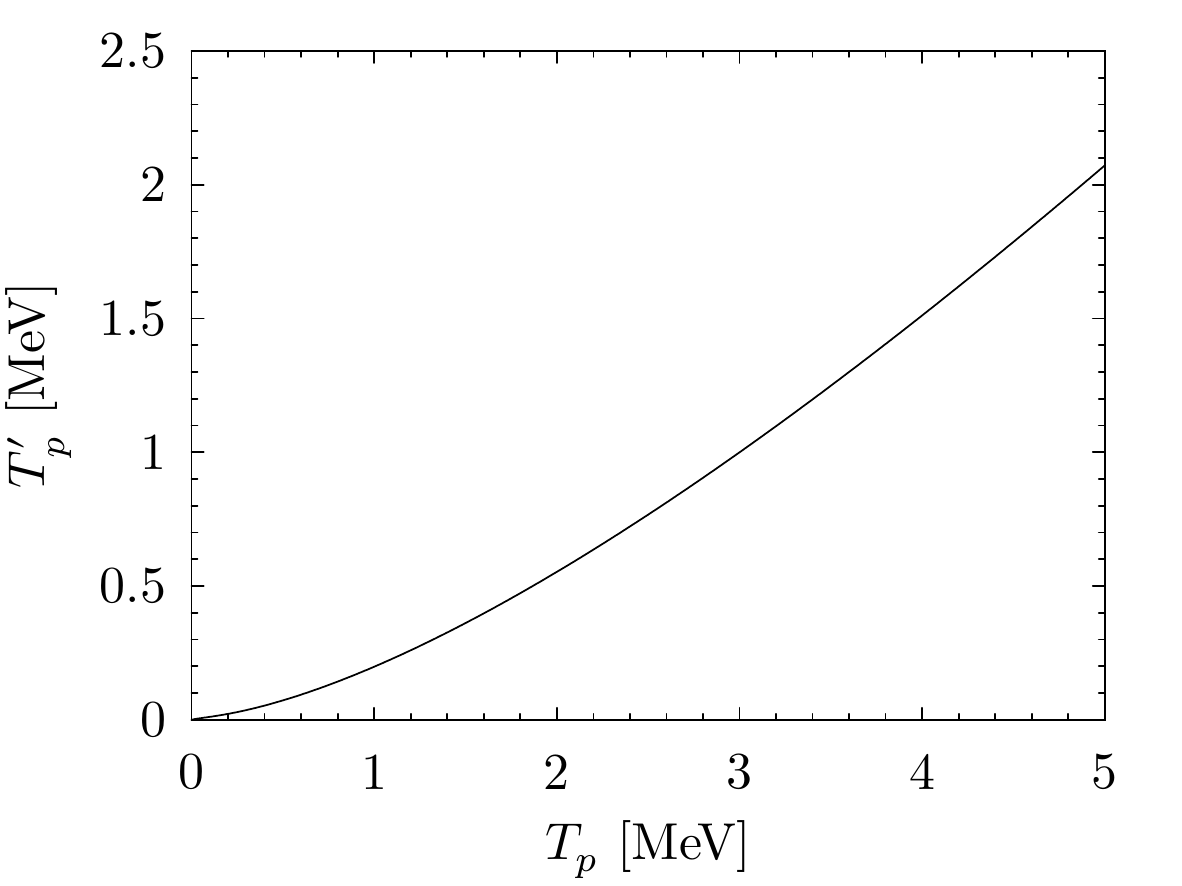}
\caption{The proton kinetic energy after quenching, $T_p'$, as a function of that before quenching, $T_p$, used in this work. The relation between $T_p'$ and $T_p$ is given in Eq.~\eqref{eq:Tp'(Tp)}.}\label{fig:quench}
\end{figure}

The differential cross section as a function of neutrino energy and the recoil energy of the final-state proton is~\cite{Beacom:2002hs}:
\begin{equation}
    \frac{d\sigma_{\nu p}(E_\nu)}{dT_p} = \frac{G_\text{F}^2 m_p}{2\pi E_\nu^2}\left[
    (c_V+c_A)^2 E_\nu^2+(c_V-c_A)^2(E_\nu-T_p)^2-(c_V^2-c_A^2)m_p T_p
    \right]
\end{equation}
where the coupling constant $c_A$ is determined by beta-decay of neutron as 0.635, and $c_V=(1-4\sin^2\theta_W)/2=0.04$. Then the cross section can be approximated in the leading order as follows:
\begin{equation}
\frac{d\sigma_{\nu p}(E_\nu)}{dT_p}\approx 4.83\times10^{-42} \text{cm}^2\cdot\text{MeV}^{-1}\vspace{0.15cm}
\times\left[ 
1+466\left(\frac{T_p}{\text{MeV}}\right)\times \left(\frac{\text{MeV}}{E_\nu}\right)^2
\right].
\end{equation}
The low-energy protons lose energy quickly by ionizations. Governed by the Bethe-Bloch stopping power formula, a 10 MeV proton will stop moving in liquid scintillator in less than $\sim0.1$ cm as we have $dE/dx\approx-100$ MeV/(g/cm$^2)$ for the proton at a few MeV. The recoil energy will be greatly quenched in the liquid-scintillator detector. The Birks' law indicates that the kinetic energy of proton after quenching is approximated by
\begin{equation}\label{eq:Tp'(Tp)}
T'_p(T_p)\approx \int^{T_p}_{0} \frac{dE}{1+k_\text{B}\langle dE/dx \rangle},
\end{equation} 
where the Birks' constant is taken $k_\text{B}\approx 0.01 ~\text{cm}/\text{MeV}$. 
The energy-loss rate of protons in the material $\langle dE/dx \rangle$ is taken from the website \textbf{PSTAR}~\cite{nist_pstar_database}.
The quenching factor (relation between $T_p$ and $T'_p$) adopted in this work is shown in Fig.~\ref{fig:quench}.

\subsection{ArCC events}
The liquid argon time projection chamber (LAr TPC) can detect SN neutrinos via their CC interactions with $^{40}$Ar. Ionization charge is drifted by electric field and collected by wire planes. We can reconstruct the 3D tracks based on wire-plane readout and the electron drifting speed, and identify particles by means of energy loss along the track. Liquid argon can act as the scintillator by itself. The first hit time and collected charge in light sensors will also help with vertex reconstructions. Therefore, LAr TPC has good angular and energy resolution with the fine wire structure while it can maintain a low threshold. We follow the description of the 40 kton-LAr TPC in DUNE~\cite{Acciarri:2015uup} focusing on the CC interaction $\nu_e+\leftidx{^{40}}{\text{Ar}}\rightarrow e^-+\leftidx{^{40}}{K^*}$, according to Ref.~\cite{Nikrant:2017nya}. 
The corresponding event number is given by
\begin{equation}\label{eq:N_Arnue}
N_{\nu_e\text{Ar}}=N_{\text{Ar}}\int^{E_\text{max}}_{E_\text{th}}\Phi(E_\nu)\sigma_{\text{Ar}\nu_e}(E_\nu)d E,
\end{equation}
where we assume $E_\text{th}=5$ MeV and $E_\text{max}=100$ MeV, $\Phi(E_\nu)$ is the fluence for $\nu_e$, $\sigma_{\nu_e\text{Ar}}$ is the cross section for $\nu_e+\leftidx{^{40}}{\text{Ar}}\rightarrow e^-+\leftidx{^{40}}{K^*}$, and $N_{\text{Ar}}\approx5.47\times10^{32}$ is the total number of $^{40}$Ar atoms in a 40 kton-LArTPC detector. 
%

\section{Constraining $\sin^22\theta_{14}$ and $\Delta m_{41}^2$ with future SN neutrino detection}\label{sec:result}

In Sec.~\ref{sec:eventnumber}, we first compute the expected event numbers from the three different types of interactions (ArCC, eES, and pES) in DUNE, Hyper-K, and JUNO detectors with our default fluence model introduced in Sec.~\ref{sec:2} for a SN occurring at $D=10$~kpc. We then derive the corresponding constraints on $\nu_e$--$\nu_s$  mixing parameters, and compare with bounds obtained by terrestrial experiments as well as from the CMB observation.
Further discussions on the dependence of our results on the SN distance, the fluence parameters, and the systematic uncertainties associated with detectors are given in Sec.~\ref{sec:sndistance}, Sec.~\ref{sec:fluuncer}, and Sec.~\ref{sec:systemerror}, respectively. 
Last, in Sec.~\ref{sec:discovery}, we comment on the sensitivity of using SN neutrino detection to discover sterile neutrinos, if they do exist.

\subsection{Event numbers and constraints with the default model}
\label{sec:eventnumber}

We combine Eq.~\eqref{eq:fluence_1} using the fluence parameters given in Table.~\ref{tab:fluence_para}, together with Eqs.~\eqref{eq:Fe_normal} or Eqs.~\eqref{eq:Fe_normal_3nu} to compute the fluence of each neutrino flavor at Earth for cases with or without $\nu_e$--$\nu_s$ mixing.
We then use Eqs.~\eqref{eq:N_eES},~\eqref{eq:N_pES}, and~\eqref{eq:N_Arnue} to compute the detected events in each detection channel listed in Table.~\ref{tab:exp_event} during the first 10~ms of the neutronization burst.
The resulting event numbers using different values of $\Delta m_{41}^2$ and $\sin^2 2\theta_{14}$ are given in Table~\ref{tab:total_rate_NO} and \ref{tab:total_rate_IO} for NO and IO, respectively.

For cases without $\nu_e$--$\nu_s$ mixing, one expects to detect more events in channels sensitive to the CC interactions (ArCC, eES) in the IO scenario than in the NO, due to the higher $P(\nu_e\rightarrow\nu_e)$ in the former (see Sec.~\ref{sec:nuesconv}).
Meanwhile, the pES channel in JUNO is only sensitive to the NC scatterings such that the expected events is independent of the mass ordering.
For NO, the expected eES event number in Hyper-K (36.5) is about a factor of 3 larger than the ArCC events (12.8) in DUNE and the sum of eES and pES events (11.3) in JUNO, due to its exceedingly large size. 
As for IO, the corresponding numbers increase to 65.3 and 41.9 for the eES in Hyper-K and ArCC in DUNE, respectively.

When we consider the $\nu_e$--$\nu_s$ mixing, one sees that for cases with small $\sin^2 2\theta_{14}$ and $\Delta m_{41}^2$ where the $\nu_e\rightarrow\nu_s$ is nonadiabatic, the expected event numbers stay the same as in the 3$\nu$ mixing scenario for both NO and IO. 
However, when $\nu_e\rightarrow\nu_s$ is adiabatic inside the SN with large $\sin^2 2\theta_{14}$ and $\Delta m_{41}^2$, the expected event numbers are reduced substantially by $\sim 60-80\%$ due to the much smaller fluence $F_{\nu_e}$ (see Eq.~\eqref{eq:Fe_normal}).
The only exception is the ArCC events of DUNE in NO, for which the reduction is minimal because $F_{\nu_e}$ is always dominated by the original $\nu_x$ flavor, $F^0_{\nu_x}$ regardless of the outcome of $\nu_e$--$\nu_s$ flavor conversion, as discussed in Sec.~\ref{sec:nuesconv}. 

\begin{table*}[b!]
\centering
\caption{\label{tab:total_rate_NO}Total event numbers for the normal ordering. The mixing angles $\theta_{12}$, $\theta_{13}$, and $\theta_{23}$ used are $33.48^\circ$, $8.5^\circ$, and $45^\circ$, respectively. The active-sterile mixing angles $\theta_{24}$ and $\theta_{34}$ are fixed at $0$.}
\begin{tabular}{|c||c|c|c|c|}
\hline\hline
$\nu_e$; @$10$~kpc (NO) & DUNE ArCC & Hyper K $e$ES  & JUNO $e$ES & JUNO $p$ES \\\hline\hline
$3$-$\nu$ mixing  & $12.8$ & $36.5$ & $2.2$ & $9.1$ \\\hline\hline
$\sin^22\theta_{14}=10^{-9}$, $\Delta m_{41}^2=10^2$ eV$^2$ & $12.8$ & $36.2$ & $2.2$ & $9$ \\\hline
$\sin^22\theta_{14}=10^{-7}$, $\Delta m_{41}^2=10^2$ eV$^2$ & $12.1$ & $27.2$ & $1.7$ & $7.7$ \\\hline
$\sin^22\theta_{14}=10^{-5}$, $\Delta m_{41}^2=10^2$ eV$^2$ & $10.2$ & $11.3$ & $0.7$ & $3.3$ \\\hline
$\sin^22\theta_{14}=10^{-3}$, $\Delta m_{41}^2=10^2$ eV$^2$ & $10.3$ & $11.3$ & $0.7$ & $3.3$ \\\hline\hline
$\sin^22\theta_{14}=10^{-9}$, $\Delta m_{41}^2=1$ eV$^2$ & $12.8$ & $36.3$ & $2.2$ & $9$ \\\hline
$\sin^22\theta_{14}=10^{-7}$, $\Delta m_{41}^2=1$ eV$^2$ & $12.7$ & $35.4$ & $2.1$ & $8.9$ \\\hline
$\sin^22\theta_{14}=10^{-5}$, $\Delta m_{41}^2=1$ eV$^2$ & $10.4$ & $12.2$ & $0.7$ & $3.9$ \\\hline
$\sin^22\theta_{14}=10^{-3}$, $\Delta m_{41}^2=1$ eV$^2$ & $10.3$ & $11.3$ & $0.7$ & $3.3$ \\\hline\hline
$\sin^22\theta_{14}=10^{-9}$, $\Delta m_{41}^2=10^{-2}$ eV$^2$ & $12.8$ & $36.3$ & $2.2$ & $9$ \\\hline
$\sin^22\theta_{14}=10^{-7}$, $\Delta m_{41}^2=10^{-2}$ eV$^2$ & $12.8$ & $36.2$ & $2.2$ & $9$ \\\hline
$\sin^22\theta_{14}=10^{-5}$, $\Delta m_{41}^2=10^{-2}$ eV$^2$ & $12.4$ & $31.3$ & $1.9$ & $8.4$ \\\hline
$\sin^22\theta_{14}=10^{-3}$, $\Delta m_{41}^2=10^{-2}$ eV$^2$ & $10.3$ & $11.3$ & $0.7$ & $3.3$ \\\hline\hline
\end{tabular} 
\end{table*}

\begin{table*}[h!]
\centering\caption{\label{tab:total_rate_IO}Total event numbers for inverted ordering. The mixing angles $\theta_{12}$, $\theta_{13}$, and $\theta_{23}$ used are $33.48^\circ$, $8.5^\circ$, and $45^\circ$, respectively. The active-sterile mixing angles $\theta_{24}$ and $\theta_{34}$ are fixed at $0$.}
\begin{tabular}{|c||c|c|c|c|}
\hline\hline
$\nu_e$; @$10$~kpc (IO) & DUNE ArCC & Hyper K $e$ES  & JUNO $e$ES & JUNO $p$ES \\\hline\hline
$3$-$\nu$ mixing & $41.9$ & $65.3$ & $3.9$ & $9.1$ \\\hline\hline
$\sin^22\theta_{14}=10^{-9}$, $\Delta m_{41}^2=10^3$ eV$^2$ & $41.8$ & $64.8$ & $3.9$ & $8.9$ \\\hline
$\sin^22\theta_{14}=10^{-7}$, $\Delta m_{41}^2=10^3$ eV$^2$ & $32.1$ & $45.2$ & $2.7$ & $7.6$ \\\hline
$\sin^22\theta_{14}=10^{-5}$, $\Delta m_{41}^2=10^3$ eV$^2$ & $7.3$ & $10.5$ & $0.6$ & $3.3$ \\\hline
$\sin^22\theta_{14}=10^{-3}$, $\Delta m_{41}^2=10^3$ eV$^2$ & $7.3$ & $10.6$ & $0.6$ & $3.3$ \\\hline\hline
$\sin^22\theta_{14}=10^{-9}$, $\Delta m_{41}^2=1$ eV$^2$ & $41.9$ & $65$ & $3.9$ & $8.9$ \\\hline
$\sin^22\theta_{14}=10^{-7}$, $\Delta m_{41}^2=1$ eV$^2$ & $40.8$ & $63$ & $3.8$ & $8.8$ \\\hline
$\sin^22\theta_{14}=10^{-5}$, $\Delta m_{41}^2=1$ eV$^2$ & $9.2$ & $12.4$ & $3.8$ & $3.8$ \\\hline
$\sin^22\theta_{14}=10^{-3}$, $\Delta m_{41}^2=1$ eV$^2$ & $7.3$ & $10.6$ & $0.6$ & $3.3$ \\\hline\hline
$\sin^22\theta_{14}=10^{-9}$, $\Delta m_{41}^2=10^{-2}$ eV$^2$ & $41.9$ & $65$ & $3.9$ & $8.9$ \\\hline
$\sin^22\theta_{14}=10^{-7}$, $\Delta m_{41}^2=10^{-2}$ eV$^2$ & $41.9$ & $56$ & $3.9$ & $9$ \\\hline
$\sin^22\theta_{14}=10^{-5}$, $\Delta m_{41}^2=10^{-2}$ eV$^2$ & $37$ & $54.2$ & $3.2$ & $8.3$ \\\hline
$\sin^22\theta_{14}=10^{-3}$, $\Delta m_{41}^2=10^{-2}$ eV$^2$ & $7.3$ & $10.6$ & $0.6$ & $3.3$ \\\hline\hline
\end{tabular} 
\end{table*}

To utilize these expected event numbers to constrain the mixing parameters $\Delta m_{41}^2$ and $\sin^22\theta_{14}$, we
define the statistics parameter $\Delta \chi^2(\sin^22\theta_{14},~\Delta m_{41}^2)$, assuming that the detected event numbers follow the Poission distributions
\begin{equation}\label{eq:chi}
\Delta\chi^2(\sin^22\theta_{14},\Delta m_{41}^2)= 2\sum_{\mathrm{event}}\left(\eta_\mathrm{event}-n_\mathrm{event}+n_\mathrm{event}\ln\frac{n_\mathrm{event}}{\eta_\mathrm{event}}\right),
\end{equation}
where $n$ and $\eta(\theta_{14}, \Delta m_{41}^2)$ denote the true total event numbers for the standard $3$-$\nu$ mixing and that predicted with $\nu_e$-$\nu_s$ mixing hypothesis, respectively. 
The subscript $_{\mathrm{event}}$ denotes a detection type in a specific detector (ArCC events in DUNE, $e$ES events in Hyper-K, $e$ES or $p$ES events in JUNO), 
and $\sigma_{\mathrm{event}}=\sqrt{n_\mathrm{event}}$.
In following analyses, we assume that the mass ordering in active neutrino sector is known, as the ordering is expected to be determined by independent terrestrial measurements with high C.L., \textit{e.g.}~ JUNO reading the shape difference in reactor neutrino spectra, and DUNE measuring matter effects in $\nu_\mu\rightarrow\nu_e$ and $\bar{\nu}_\mu\rightarrow\bar{\nu}_e$ oscillations.

\begin{figure}[h!]
\centering
\includegraphics[width=0.48\textwidth]{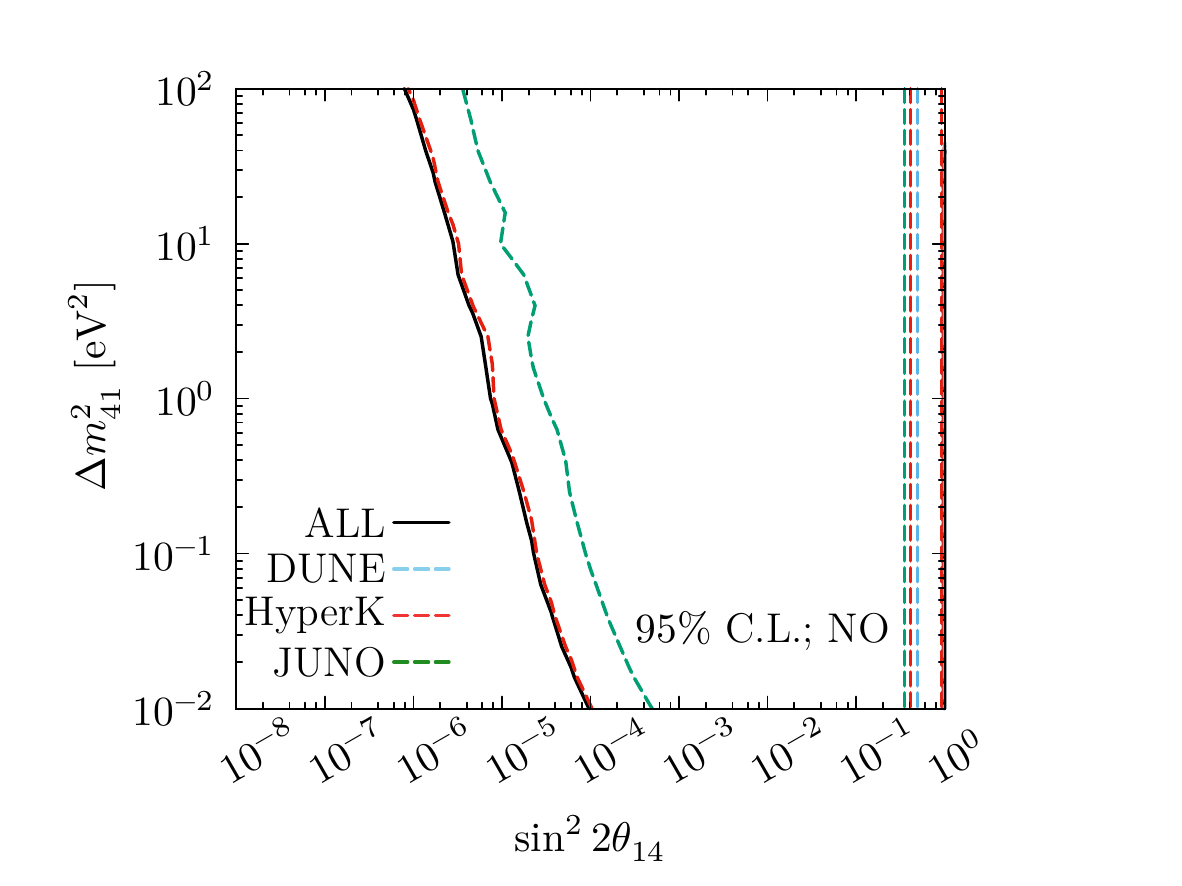}\hspace{-1.2cm}
\includegraphics[width=0.48\textwidth]{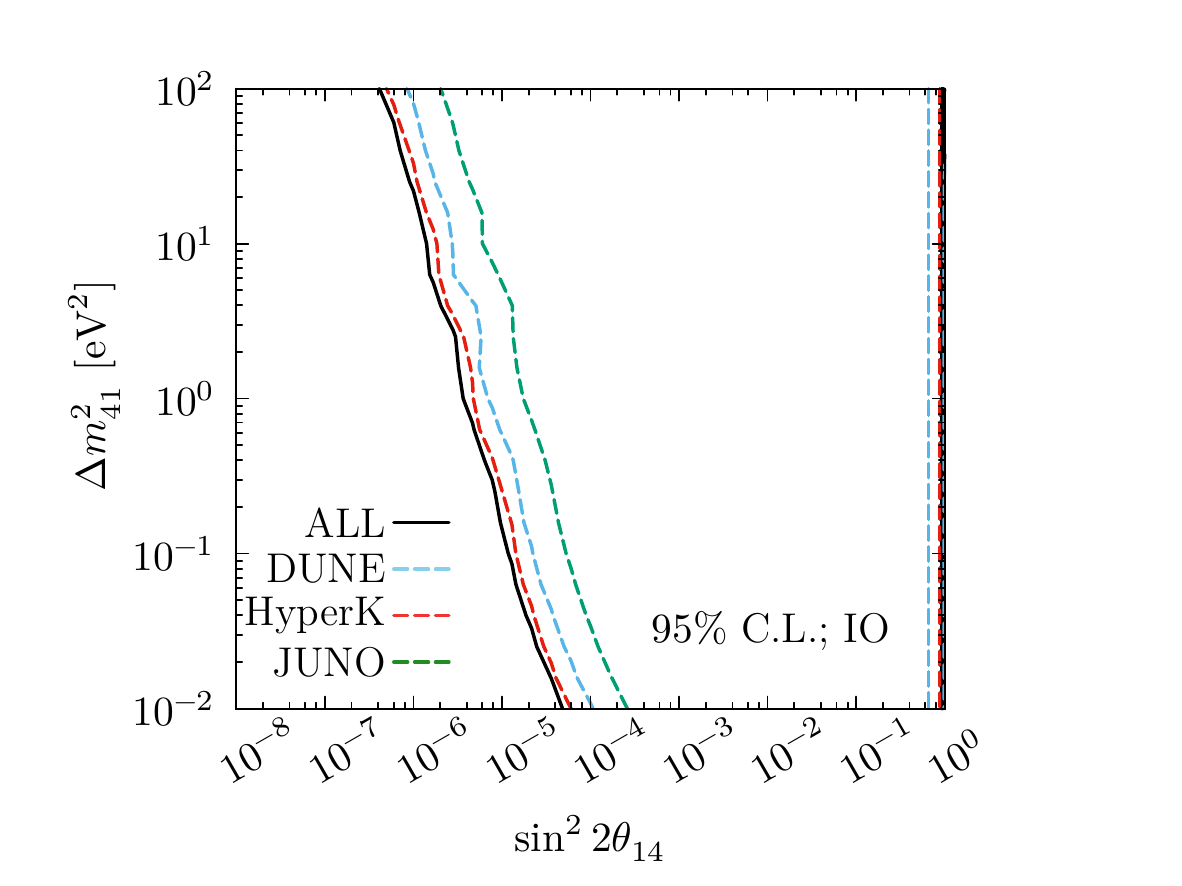}
\caption{The exclusion contours at $95\%$ C.L. on the $\sin^22\theta_{14}$-- $\Delta m^2_{41}$ plane for a SN occuring at $10$ kpc 
for the normal (left) and inverted (right) ordering. Considered cases are DUNE (dashed-blue), Hyper-K (dashed-red), JUNO (dashed green), and the combination of them (solid-black). 
}
              \label{Fig:Constraint_exp}%
\end{figure}

We show the exclusion limits at $95\%$ C.L. on the $\sin^22\theta_{14}$--$\Delta m_{41}^2$ plane for DUNE (ArCC events), Hyper-K ($e$ES events), and JUNO ($e$ES and $p$ES events) separately, as well as that result when all events from these detectors are combined, assuming a SN occurring at $10$ kpc. We find that the dominating contribution to the total sensitivity comes from the reduction of $e$ES events in Hyper-K in both NO and IO. JUNO alone can also provide constraints for either mass ordering that are a bit less stringent than Hyper-K. 
As for DUNE, it can yield comparable constraints as from Hyper-K for IO. 
However, the sensitivity disappears in NO as the ArCC events are not affected by $\nu_e$-$\nu_s$ flavor conversion in most of the parameter space, consistent with what we conclude in Tables~\ref{tab:total_rate_NO} and \ref{tab:total_rate_IO}.

\begin{figure}[h!]
\centering
\includegraphics[width=0.48\textwidth]{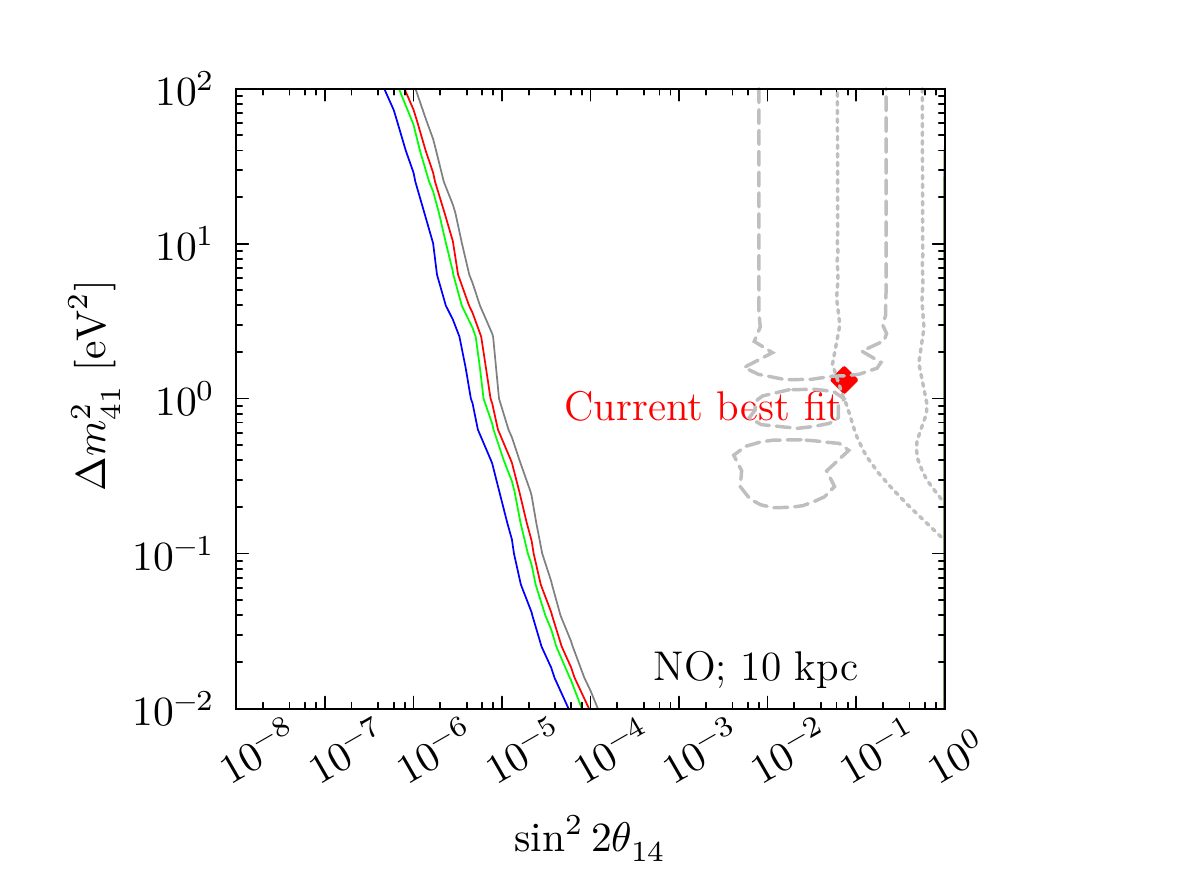}\hspace{-1.2cm}
\includegraphics[width=0.48\textwidth]{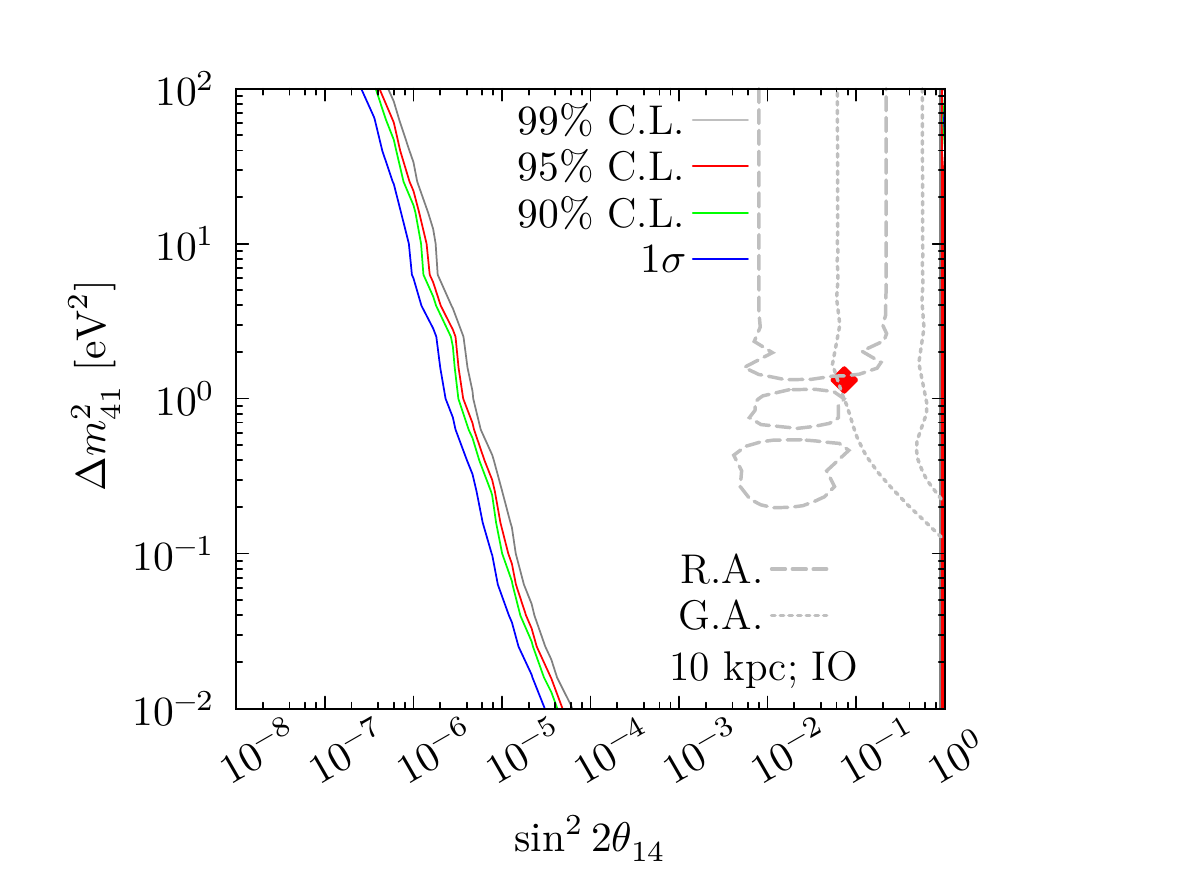}
\caption{The exclusion contours at $1\sigma$, $90\%$, $95\%$, and $99\%$ C.L. on the $\sin^22\theta_{14}$-- $\Delta m^2_{41}$ plane for a SN occuring at $10$ kpc 
for the normal (left) and inverted (right) ordering. 
The red diamond indicates the best-fit values taken from Ref.~\cite{Diaz:2019fwt}.
The gray dashed and dotted curves are from the reactor anomaly (R.A.)~\cite{Mention:2011rk} and gallium anomaly (G.A.)~\cite{Giunti:2010zu}.}
              \label{Fig:Constraint_CL}%
\end{figure}

Fig.~\ref{Fig:Constraint_CL} shows the exclusion limits at $1\sigma$, $90\%$, $95\%$ and $99\%$ confidence level (C.L.) derived using Eq.~\eqref{eq:chi} with our default model assumptions for the NO (left panel) and IO (right panel) on the ($\Delta m_{41}^2$, $\sin^22\theta_{14}$) plane.
The best-fit values combining various experiments taken from Ref.~\cite{Diaz:2019fwt}, $\sin^22\theta_{14}=0.053$, $\Delta m_{41}^2=1.32$ eV$^2$, are indicated by the red diamond on the same plots.
Also shown are the allowed parameter space from the reactor anomaly~\cite{Mention:2011rk} and Gallium anomaly~\cite{Giunti:2010zu}.
Clearly, regions with $\sin^2 2\theta_{14}\gtrsim 10^{-5}\times\sqrt{{\rm 1~eV^2}/\Delta m^2_{41}}$, including the best-fit region hinted by the terrestrial experiments, can be robustly tested by the detection of the SN neutrino signals during the neutronization burst phase for a SN occurring at 10~kpc, independent of the mass ordering.
Comparing the exclusion limits in NO and IO, the excluded regions slightly extend to smaller $\sin^2 2\theta$ for IO as the expected event numbers without $\nu_e$--$\nu_s$ mixing are larger.
For bounds at large $\sin^22\theta_{14}$, they can only be found in IO, but not in NO.
This is because the constraints here are derived mainly due to changes in $P(\nu_e\rightarrow\nu_e)$ (see Fig.~\ref{Fig:prob}), and the DUNE event numbers in IO are much larger than in NO (see Table.~\ref{tab:total_rate_NO} and \ref{tab:total_rate_IO}).
We note that a different choice of $t_1=20$~ms, integrating over a longer-time window for the fluence (see Sec.~\ref{sec:SNnu}), leads to nearly identical exclusion limits as in Fig.~\ref{Fig:Constraint_CL}. As bounds for the large $\sin^22\theta_{14}$ are mainly determined by the rapid change of $P(\nu_e\rightarrow\nu_e)$ in $\sin^22\theta_{14}$, we will focus on the left bounds only in Secs.~\ref{sec:sndistance}, \ref{sec:fluuncer}, \ref{sec:systemerror}.

\begin{figure}[h!]
\centering
\includegraphics[width=0.48\textwidth]{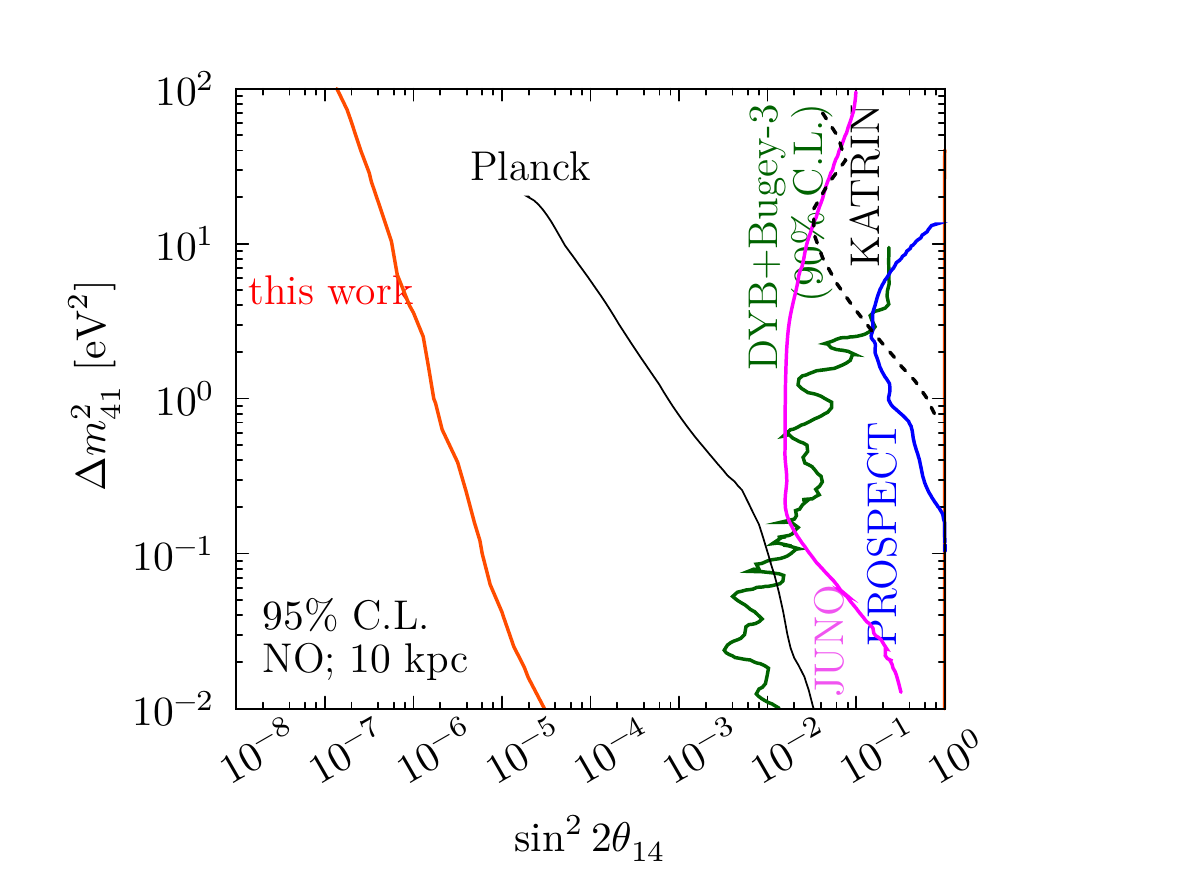}\hspace{-1.2cm}
\includegraphics[width=0.48\textwidth]{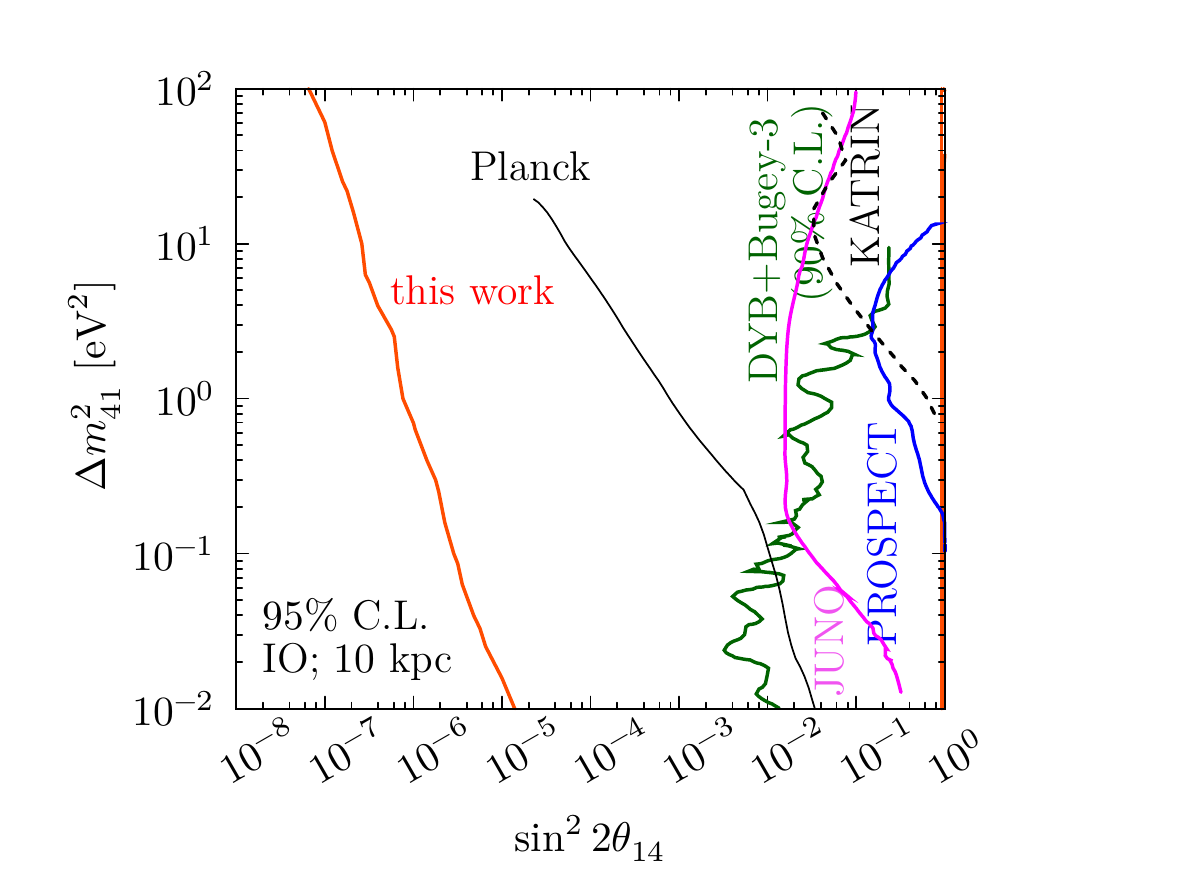}
\caption{The exclusion contours at $95\%$ C.L. on $\sin^22\theta_{14}$ and $\Delta m^2_{41}$ for a SN at $10$ kpc 
for the normal (left) and inverted (right) ordering. 
The sensitivity or constraint of other experiments are also presented. The combination of Daya Bay and Bugey-3 (green curve) is taken from Ref.~\cite{Adamson:2020jvo}. The result for Planck is taken from Ref.~\cite{Mirizzi:2013gnd}. The sensitivity of ongoing experiment PROSPECT with 33 live-days of reactor-ON data is from Ref.~\cite{Ashenfelter:2018iov}.
Also shown are the predicted sensitivity for JUNO with 450 days of data-taking from Ref.~\cite{An:2015jdp} and that for KATRIN taken from Ref.~\cite{Formaggio:2011jg}.
Except the result for the combination of Daya Bay and Bugey-3 that is at $90\%$~C.L., the others are at $95\%$~C.L.}
              \label{Fig:Constraint_compare}%
    \end{figure}
    
In Fig.~\ref{Fig:Constraint_compare}, we compare the exclusion limits at $95\%$~C.L 
to bounds obtained by current/future experiments, as well as those inferred from the CMB observation.
The red curves are the limits derived above, while the black, green, blue curves are bounds set by Planck (taken from Ref.~\cite{Mirizzi:2013gnd}), by the combination of DayaBay and Bugey-3 (taken from Ref.~\cite{Adamson:2020jvo}), and by PROSPECT (taken from Ref.~\cite{Ashenfelter:2018iov}), respectively.
Also shown by the magenta and black-dot curves are the projected sensitivity by JUNO (taken from Ref.~\cite{An:2015jdp}) and KATRIN (taken from Ref.~\cite{Formaggio:2011jg}).
Clearly, the projected limit from a future SN neutrino detection discussed here can outperform all other bounds, including the strong limit set by Planck, in the parameter space of interest.

\subsection{Dependence on the SN distance}\label{sec:sndistance}

   \begin{figure}[h!]
   \centering
   \includegraphics[width=0.48\textwidth]{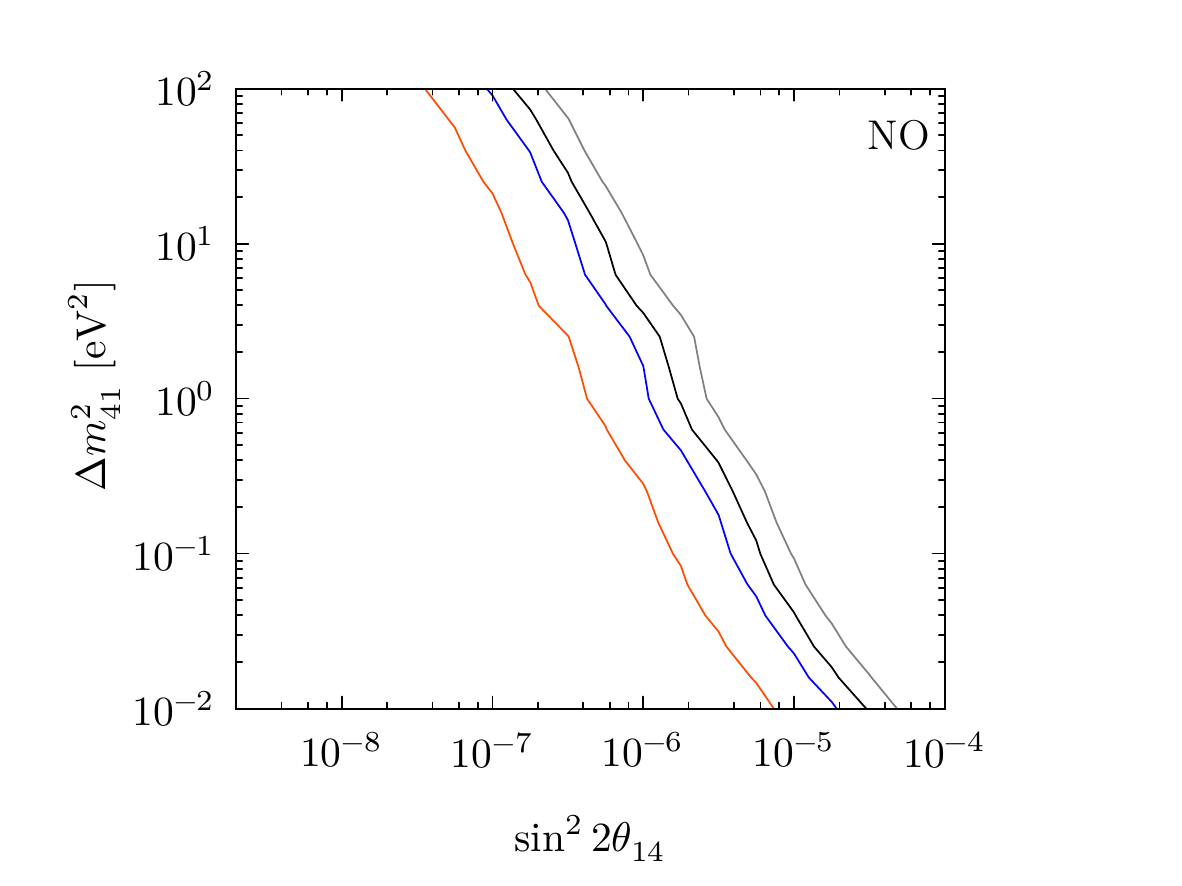}\hspace{-1.2cm}
   \includegraphics[width=0.48\textwidth]{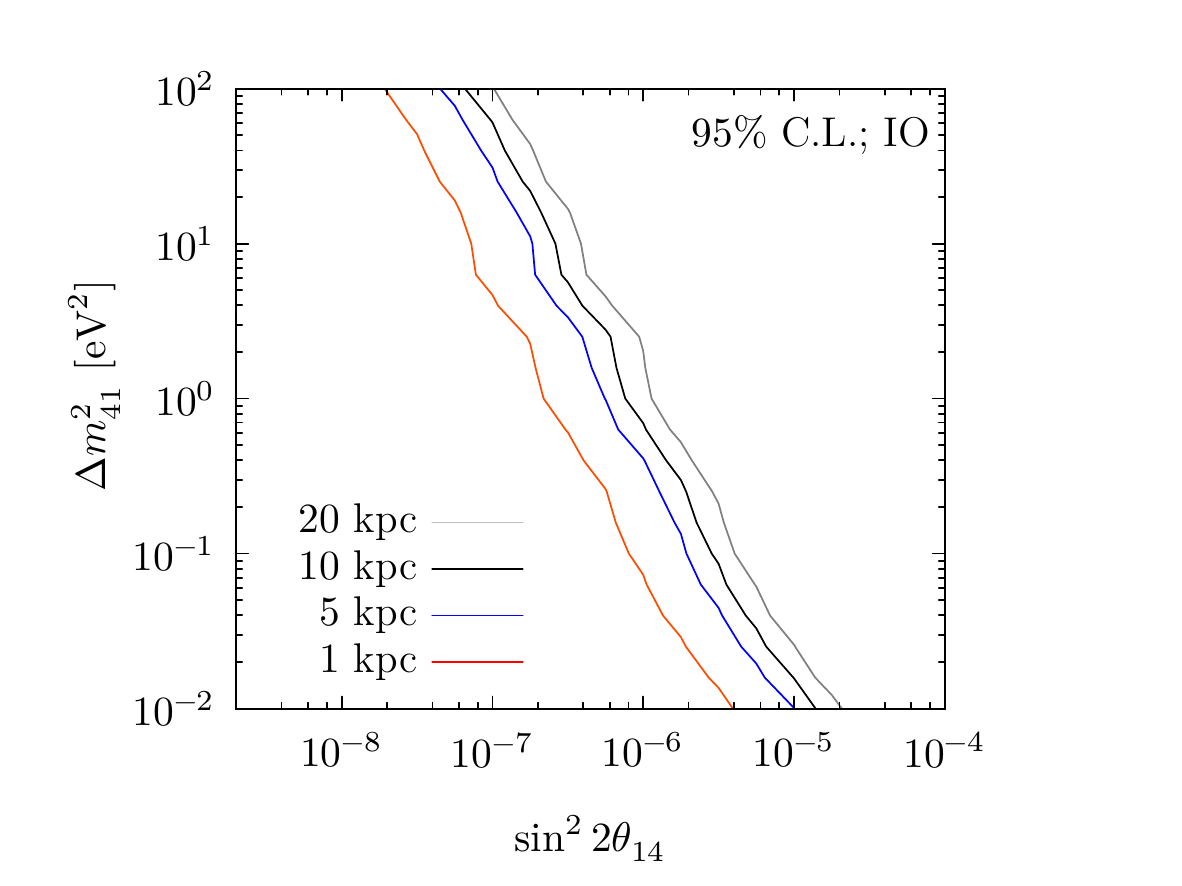}
   \caption{The exclusion contour at $95\%$ C.L. on $\sin^22\theta_{14}$ and $\Delta m^2_{41}$ for a SN at $20$, $10$, $5$, $1$ kpc, 
   for the normal (left) and inverted (right) orderings. }
              \label{Fig:Constraint_dis}%
    \end{figure}
    
As the neutrino fluence at Earth is inversely proportional to the square of the distance to the SN, $\Phi_{\nu_\alpha}\propto 1/D^2$ (see Eqs.~\ref{eq:fluence_1}), the exclusion limits obviously will be strengthened (reduced) when $D$ is smaller (larger) than 10~kpc adopted in the default model assumption.
In Fig.~\ref{Fig:Constraint_dis}, we show the exclusion contours on $\sin^22\theta_{14}$ and $\Delta m_{41}^2$ at $95\%$ C.L. for both mass orderings with $D=1,~5,~10,~20$~kpc. 
Within this distance range, the case with smaller $D$ can clearly exclude a larger range of the parameter space. 
However, we note that because $P(\nu_e\rightarrow\nu_e)$ and $P(\nu_e\rightarrow\nu_x)$ approach asymptotic values outside the transition regime from non-adiabatic to adiabatic flavor conversion, the derived exclusion limits will not further move to larger (smaller) $\sin^22\theta_{14}$ for much larger (smaller) value of $D$.

\begin{figure}
\centering
\includegraphics[width=0.48\textwidth]{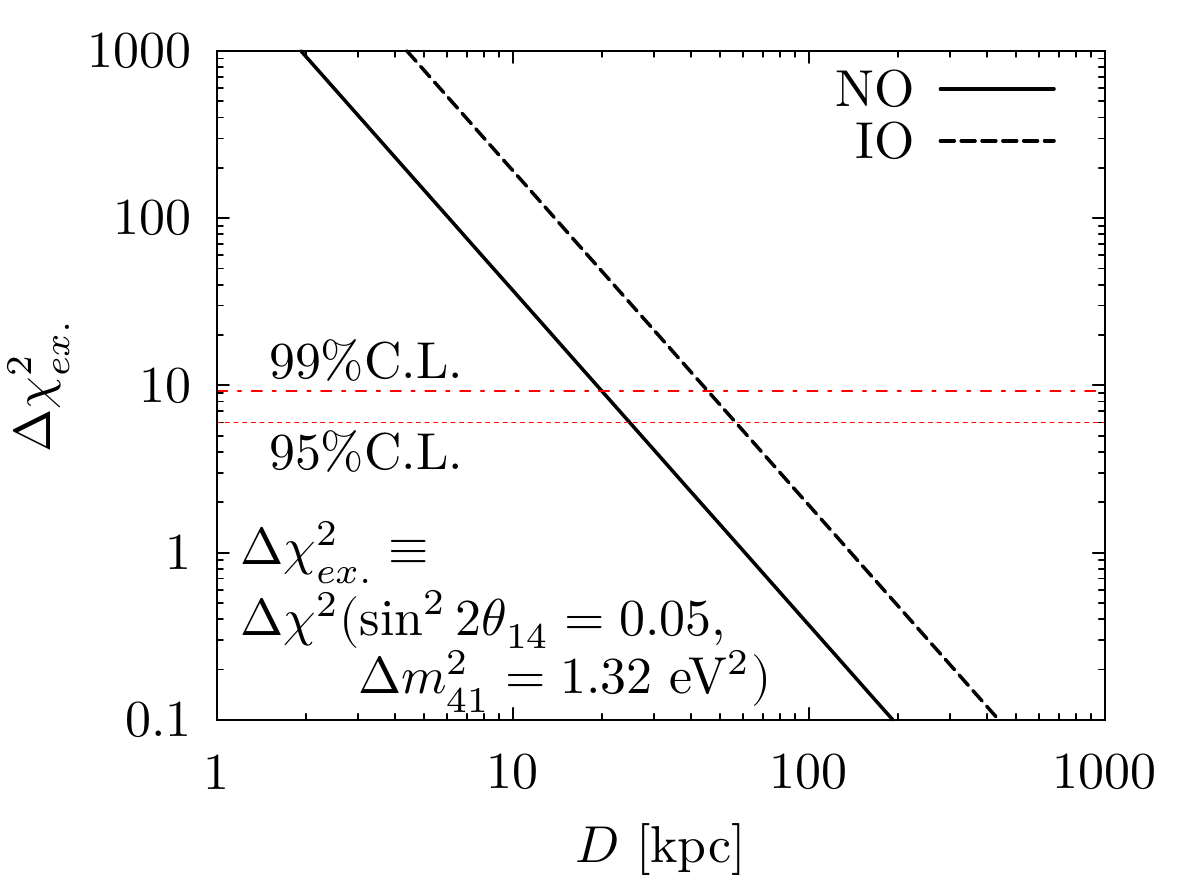}
\caption{The $\Delta \chi^2(0.05,~1.32~\mathrm{eV}^2)$ (see Eq.~\eqref{eq:chi}) as a function of the SN distance $D$, 
for the normal (solid) and inverted (dashed) ordering.}
\label{Fig:Ex_dis}%
\end{figure}
   
Fig.~\ref{Fig:Ex_dis} further shows the distance sensitivity of using SN neutronization burst events to probe the parameter region of anomalies. Taking the best-fit values $\Delta\chi^2(\sin^22\theta_{14}=0.053$, $\Delta m_{41}^2=1.32$ eV$^2)$, one can see that for a SN occuring at $D\lesssim 20 (50)$~kpc in NO (IO) , the SN neutronization burst signals can 
test if the reported $\nu_e$--$\nu_s$ mixing parameters, up to $99\%$~C.L., is compatible with the SN neutrino detection or not.

\subsection{Impact of the fluence parameters}
\label{sec:fluuncer}

\begin{figure}[h!]
\centering
\includegraphics[width=0.48\textwidth]{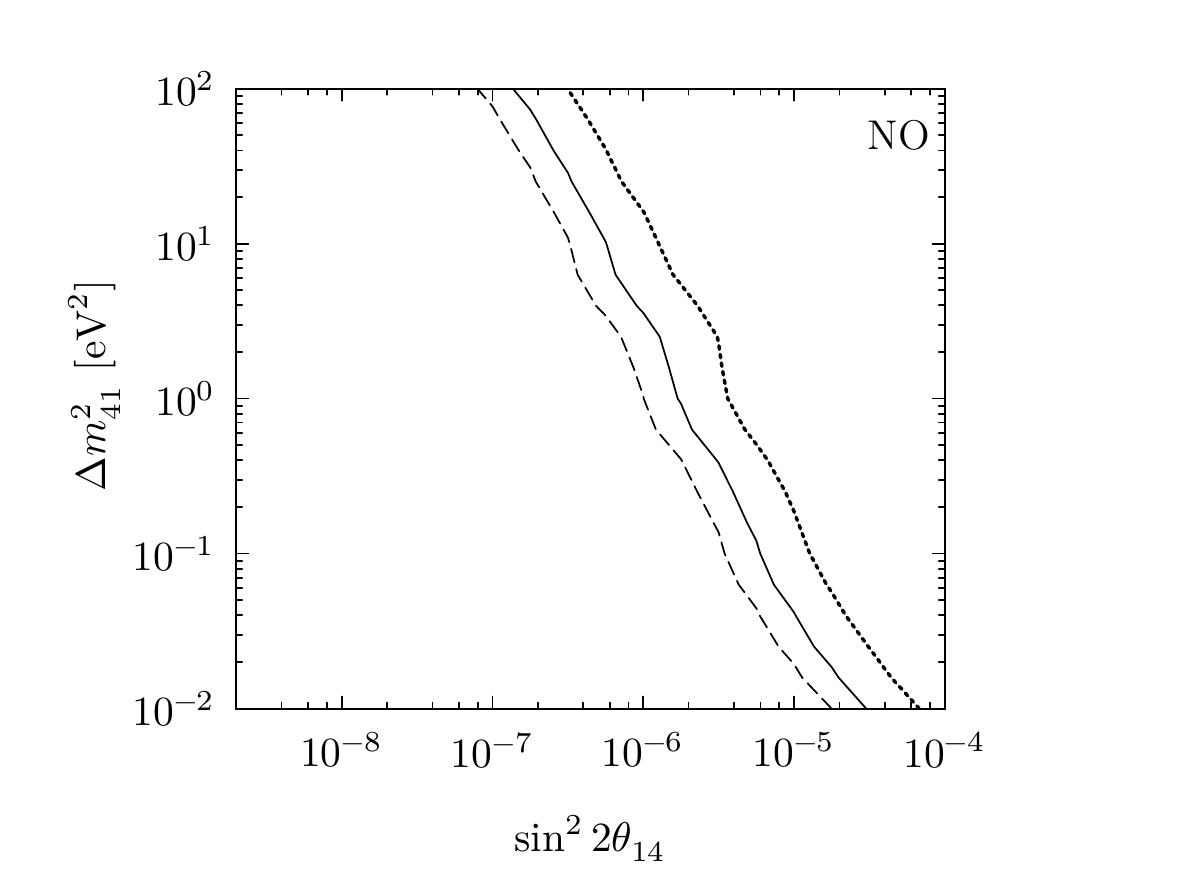}\hspace{-1.2cm}
\includegraphics[width=0.48\textwidth]{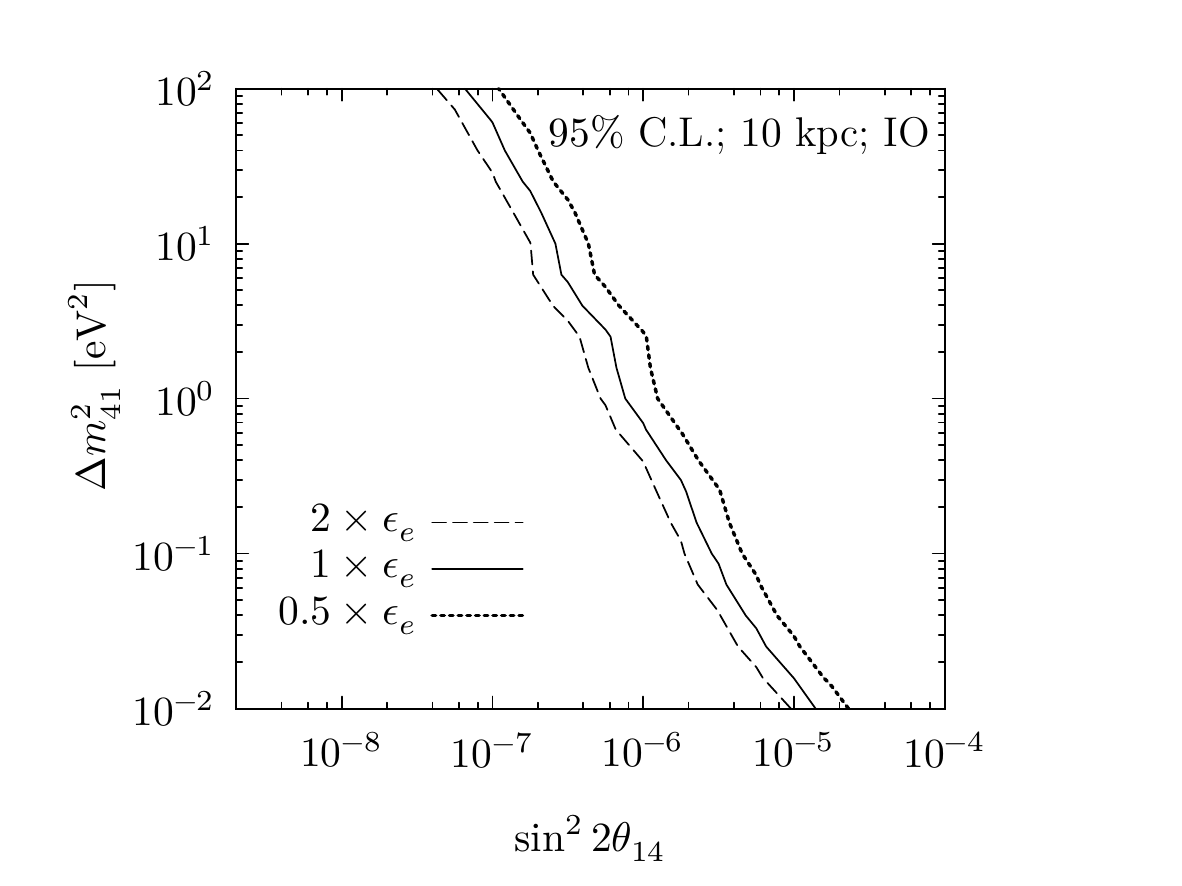}
\caption{The exclusion contours at $95\%$ C.L. for a SN at $10$ kpc,
for the normal (left) and inverted (right) ordering. 
Three values of $\epsilon_{\nu_e}$ are considered: $2\times\epsilon_{\nu_e}$ (dashed), $\epsilon_{\nu_e}$ (solid), and $0.5\times\epsilon{\nu_e}$ (dotted), where $\epsilon_{\nu_e}$ is the value in Table~\ref{tab:fluence_para}. }
\label{Fig:Constraint_flu}%
\end{figure}

Past studies showed that the prediction of neutrino luminosities and average energies for all flavors during the neutronization burst phase is relatively robust and insensitive to the assumptions of the underlying nuclear equations of states, the stellar progenitor mass, and the numerical implementation~\cite{Kachelriess:2004ds,OConnor:2018sti}.
Nevertheless, two main sources of uncertainties that enter the assumption of the neutrino fluence exist and should be considered.
First, there can be up to $\sim 50\%$ of variations in $L_{\nu_e}$ and smaller changes in $\langle E_{\nu_e}\rangle$~\cite{Kachelriess:2004ds,OConnor:2018sti}. 
Second, the rise-time of the luminoisity of other flavors can be a bit more uncertain~\cite{Serpico:2011ir,OConnor:2018sti}.
Both factors can affect the predicted event numbers in the detectors.
Thus, here we explore how our derived exclusion limits depend on these two main uncertainty sources by taking following extreme assumptions for the fluence parameters beyond uncertainties reported in the literature. 

\begin{figure*}[h!]
\centering
\includegraphics[width=0.48\textwidth]{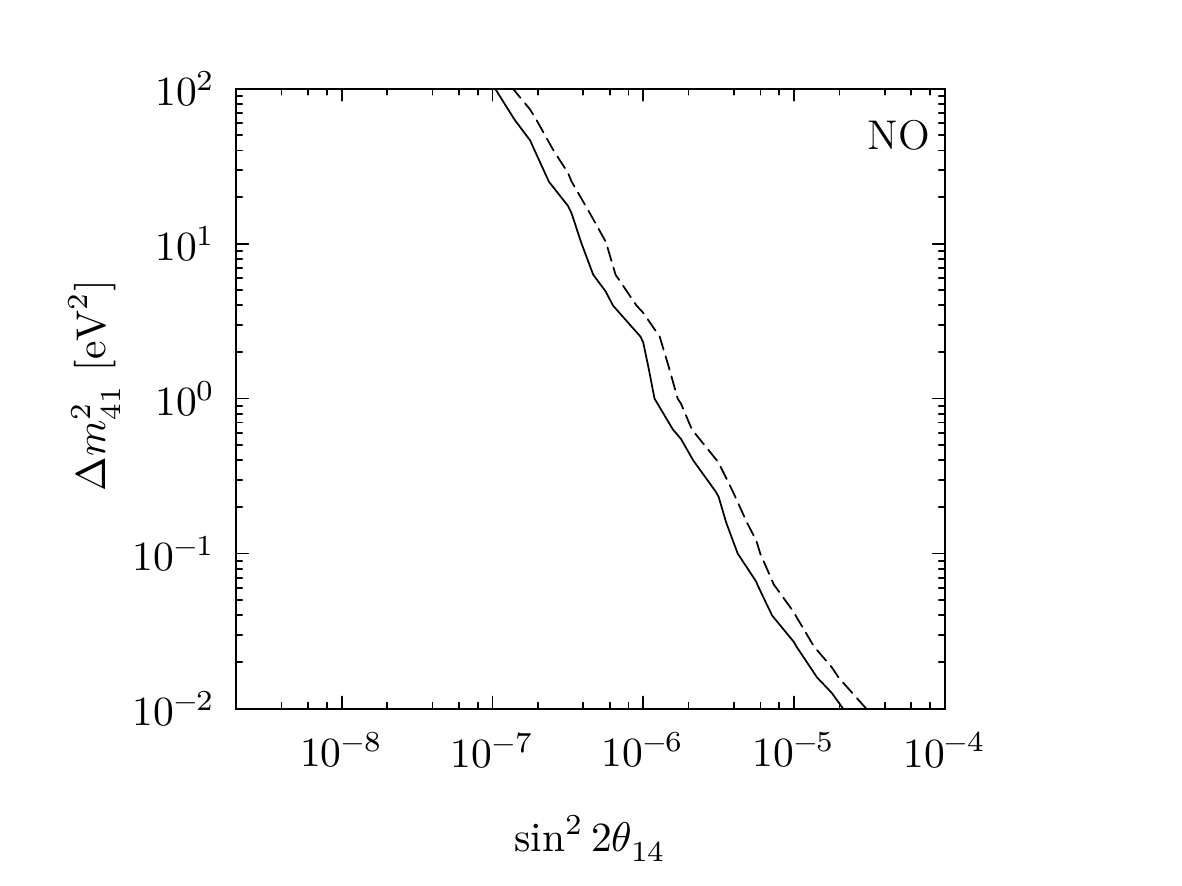}\hspace{-1.2cm}
\includegraphics[width=0.48\textwidth]{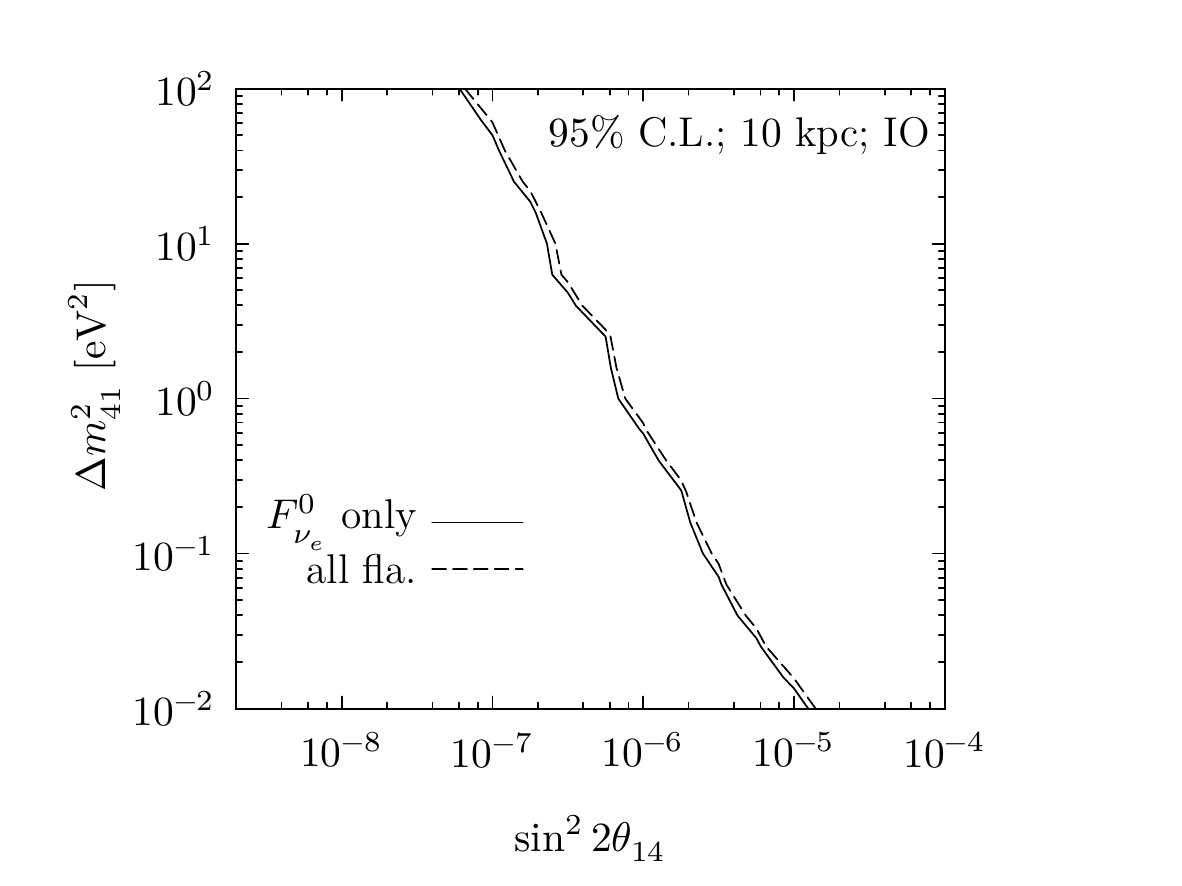}
\caption{The exclusion contour at $95\%$ C.L. on $\sin^22\theta_{14}$ and $\Delta m^2_{41}$ with CCSN $\nu$ events from $10$ kpc for the normal (left) and inverted (right) ordering. The solid lines are derived by considering only the $F_{\nu_e}^0$ ($F_{\bar\nu_e}^0=F_{\bar\nu_x}^0=0$), while the dashed ones include all flavors of neutrinos.}\label{Fig:Constraint_bg}%
\end{figure*}

We first vary the $\epsilon_{\nu_e}$ by a factor of $2$ from our default model to account for the uncertainty in the $L_{\nu_e}$ and $\langle E_{\nu_e}\rangle$ and show the results in Fig.~\ref{Fig:Constraint_flu}. 
The resulting exclusion limits are only weakened by $\lesssim 2$ in $\sin^22\theta_{14}$ for a given $\Delta m^2_{41}$ in both mass orderings\footnote{We choose to test the dependence on the fluence by taking such an extreme variation here because these uncertainties cannot be properly described by a likelihood function.}.
Furthermore, we test the impact of the uncertain rise-time in the luminosities of other neutrino flavors by setting $F_{\nu_x}=F_{\bar\nu_e}=0$. 
Fig.~\ref{Fig:Constraint_bg} shows that this has little impact on the obtained exclusion limits.
Both of these tests show that the derived bounds are rather insensitive to the fluence parameters taken in this study.

\subsection{Impact of other uncertainties}
\label{sec:systemerror}
   
   \begin{figure*}[h!]
   \centering
   \includegraphics[width=0.48\textwidth]{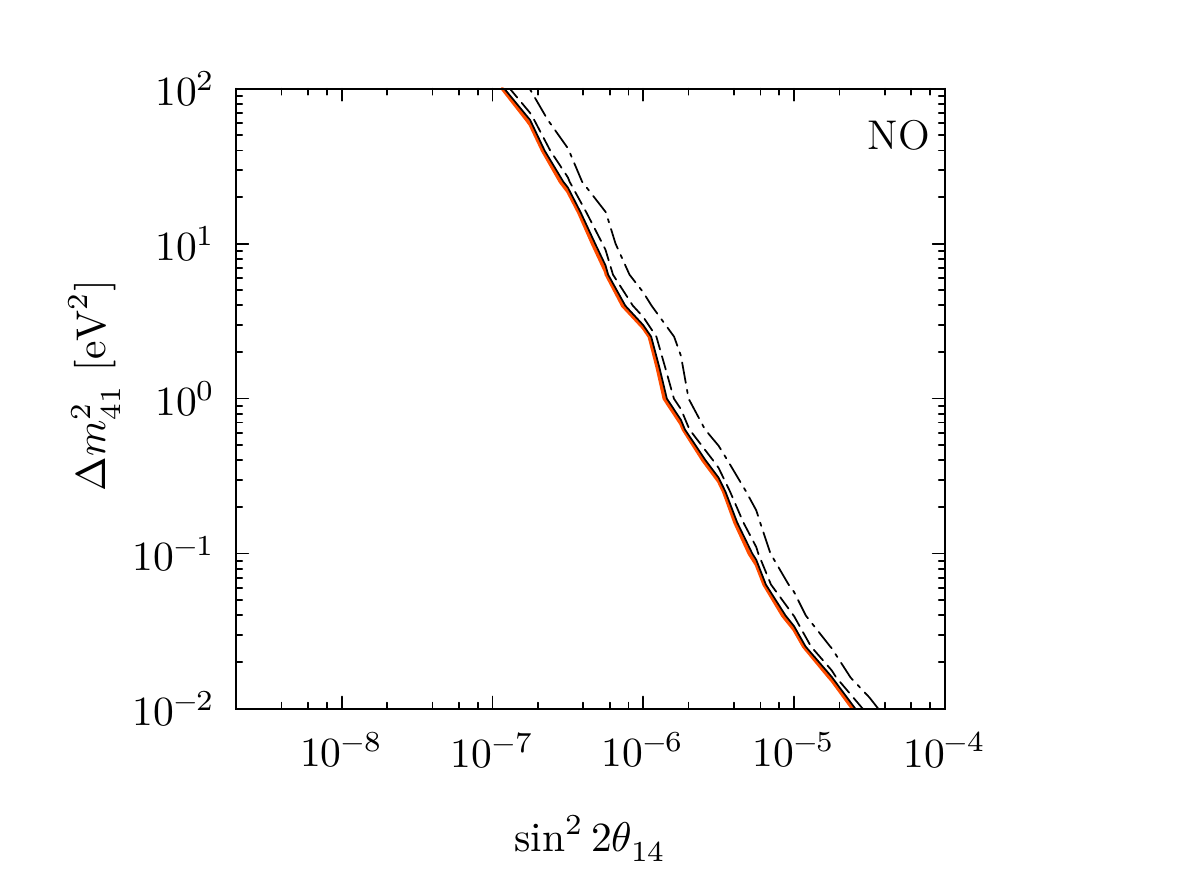}\hspace{-1.2cm}
   \includegraphics[width=0.48\textwidth]{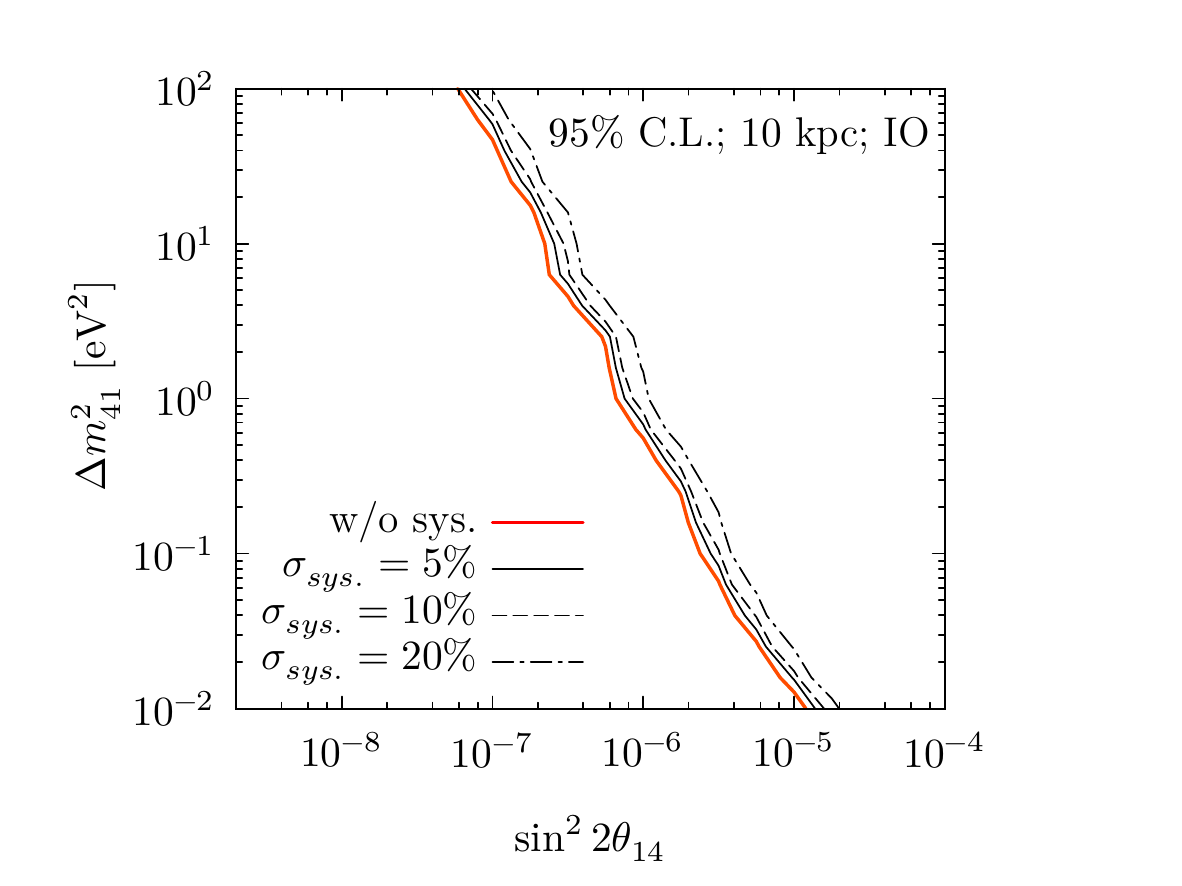}
   \caption{The exclusion contour at $95\%$ C.L. on $\sin^22\theta_{14}$ and $\Delta m^2_{41}$ with CCSN$\nu$ events from $10$ kpc for the normal (left) and inverted (right) ordering, with systematic errors described by a Gaussian distribution. Standard deviations in the Gaussian distribution with regards to detector responses are studied: $5\%$, $10\%$, $20\%$ for all experiments.}
              \label{Fig:Constraint_sys}%
    \end{figure*}

Aside from the uncertainties in the SN neutrino fluence, the systematic uncertainties in detectors can also slightly affect the derived exclusion limits.
To explore this effect, we define a new 
statistical parameter $\Delta\chi^2_{sys}$ as
\begin{equation}
\Delta\chi^2_{sys}=\min_{\mathrm{ALL}~\xi_{\mathrm{event}}}\sum_{\mathrm{event}}\Delta\chi^2_{\mathrm{event}}(\sin^22\theta_{14},~\Delta m_{41}^2,~\xi_{\mathrm{even t}}),    
\end{equation}
where
\begin{align}
\Delta\chi^2_{\mathrm{event}}(\sin^22\theta_{14},~\Delta m_{41}^2,~\xi_{\mathrm{even t}})=
&2\sum_{\mathrm{event}}\Big((1+\xi_{\mathrm{event}})\eta_\mathrm{event}-n_\mathrm{event}\\\nonumber & +n_\mathrm{event}\ln\frac{n_\mathrm{event}}{(1+\xi_{\mathrm{event}})\eta_\mathrm{event}}\Big)+\frac{\xi_{\mathrm{event}}^2}{\sigma^2_{\mathrm{event}}},
\end{align}
with $\xi_{\mathrm{event}}$ and $\sigma_{\mathrm{event}}$ being the nuisance parameter and the Gaussian width for each event, respectively. 
And the minimization "$\min_{\mathrm{ALL}~\xi_{\mathrm{event}}}$" means minimizing over all nuisance parameter $\xi_{\mathrm{event}}$. 
As in Eq.~\eqref{eq:chi}, the subscript $_\mathrm{event}$ represents an interaction type in a detector. 
Here, we assign each type an independent nuisance parameter and assume that there are no correlations between different event types.
We vary the nuisance parameters $\xi_{\mathrm{event}}$ in the Gaussian distribution with $5\%$, $10\%$, and $20\%$ variations of $\sigma_{\mathrm{event}}$ to obtain the minimum of $\Delta\chi^2_{sys}$.
As shown in Fig.~\ref{Fig:Constraint_sys}, we find that including the systematic errors up to $10\%$ has almost negligible impact for both mass orderings. Larger impacts are found in the case with $\sigma_{sys.}$ of $20\%$. In IO, the impact is similar to that with $0.5\times\epsilon_e$, shown in Fig.~\ref{Fig:Constraint_flu}, while the difference in NO is less significant.

Another error source is the uncertainty in the SN distance $D$ which may be measured independently via the measurement of electromagnetic signals or gravitational waves. 
Similarly to the effect of the above systematic uncertainties, an uncertainty of $20\%$ in distance only results in negligible changes to the exclusion limits.
%

\subsection{Discovering sterile neutrino with SN neutrinos?}
\label{sec:discovery}

   \begin{figure*}[h!]
   \centering
   \includegraphics[width=0.48\textwidth]{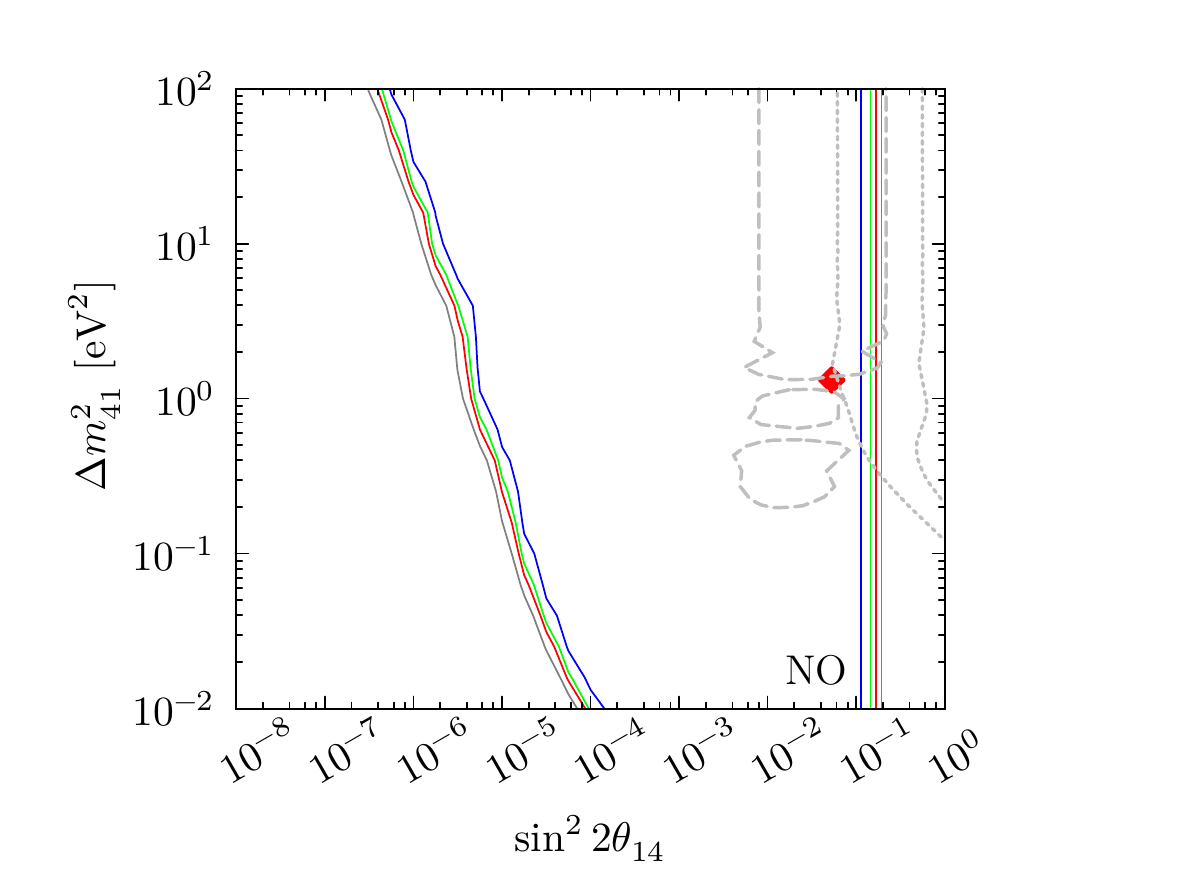}\hspace{-1.2cm}
   \includegraphics[width=0.48\textwidth]{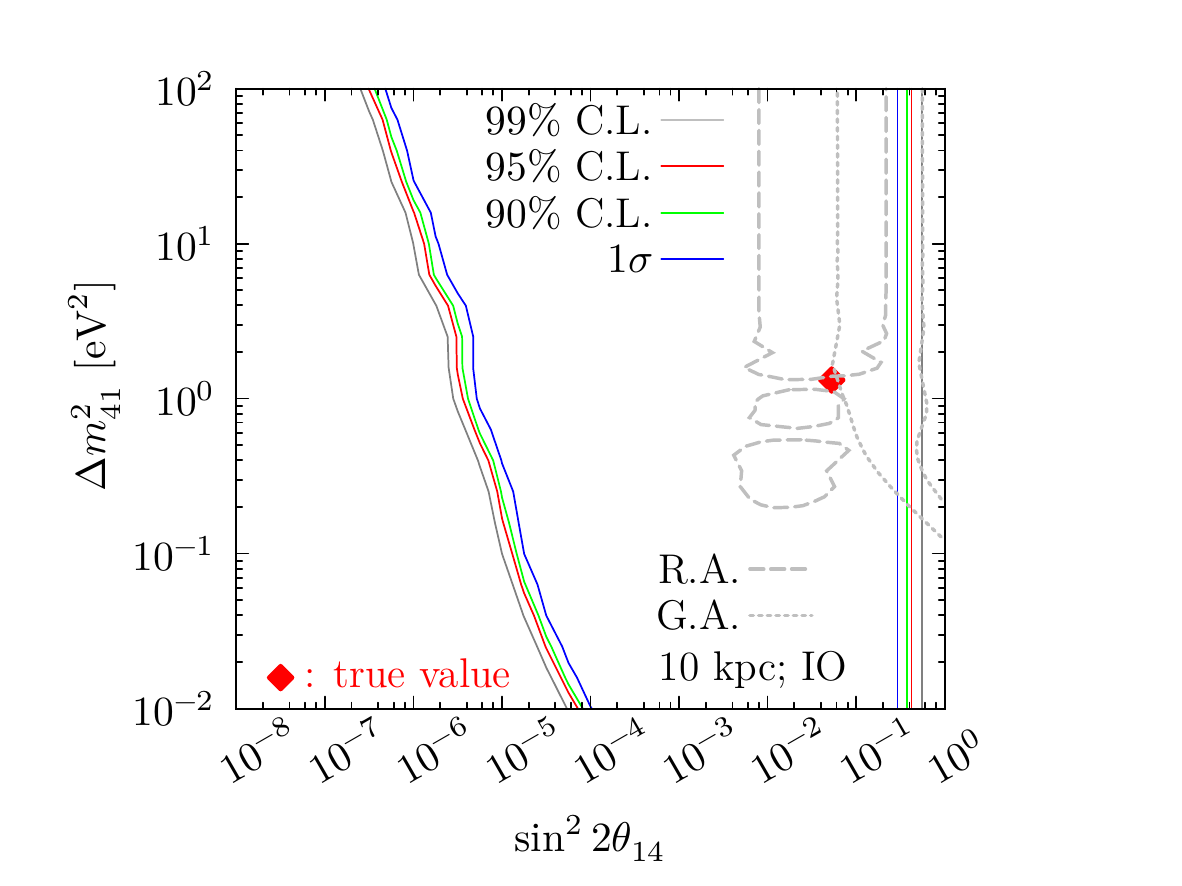}
   \caption{The exclusion contour at $1\sigma$, $90\%$, $95\%$, and $99\%$ C.L. on $\sin^22\theta_{14}$ and $\Delta m^2_{41}$ with CCSN$\nu$ events from $10$ kpc, assuming the the true values at the current best fit ($\sin^22\theta_{14}=0.053$, $\Delta m_{41}^2=1.32$~eV$^2$) for the normal (left) and inverted (right) ordering. The statistics quantity studied in this figure is defined in Eq.~\eqref{eq:chi_bf}. 
   The abbreviation "R.A." stands for the reactor anomaly by the dashed line while "G.A." is the Gallium anomaly by the dotted line.
   }
              \label{Fig:Constraint_bf}%
    \end{figure*}

Up to now, we only examine the exclusion limits on $\nu_e$--$\nu_s$ mixing using the detection of events from the SN neutronization neutrino burst. 
Another interesting question to ask is: 
If light sterile neutrinos do exist and mix with electron neutrinos, can we pin down their parameter space using imminent SN neutrino data? To address this question, we further consider a hypothetical case assuming that $\nu_e$--$\nu_s$ mixing parameters are driven by the current best-fit values with $\sin^22\theta_{14}=0.053$ and $\Delta m_{41}^2=1.32$~eV$^2$, which gives rise to a true total event number $n^{b.f.}_{\mathrm{event}}$. 
Comparing this to the event numbers predicted by other values of $\eta(\sin^22\theta_{14}, \Delta m_{41}^2)$, we then quantify the discovery potential by another statistical distribution $\Delta\chi^2_{b.f.}$,
\begin{small}
\begin{equation}\label{eq:chi_bf}
\Delta\chi^2_{b.f.}(\sin^22\theta_{14},\Delta m_{41}^2)=2\sum_{\mathrm{event}}\left(\eta_\mathrm{event}-n^{b.f.}_\mathrm{event}+n^{b.f.}_\mathrm{event}\ln\frac{n^{b.f.}_\mathrm{event}}{\eta_\mathrm{event}}\right),
\end{equation}
\end{small}
to answer the question posted above
(assuming again that the neutrino mass ordering in the active sector is known).

Fig.~\ref{Fig:Constraint_bf} shows the resulting exclusion contours at $1\sigma$, $90\%$, $95\%$, and $99\%$ C.L. in such a scenario with, once again, our default model parameters with $D=10$~kpc.
As expected, the shapes of the limits here closely resemble those derived in Sec.~\ref{sec:eventnumber}.
One can only exclude the parameter spaces with $\sin^2 2\theta_{14}\lesssim 10^{-5}\times \sqrt{1~{\rm eV}^2/\Delta m^2_{41}}$, as well as $\sin^2 2\theta_{14}\gtrsim \mathcal{O}(0.1)$, using the total events alone, because the event numbers within the contours are insensitive to the mixing parameters (see also Table.~\ref{tab:total_rate_NO} and \ref{tab:total_rate_IO}).
Note that here the NO case can probe better than the IO case at large $\sin^2 2\theta_{14}$.

Based on these results, one can conclude that although using the detected event numbers of SN neutrinos during the neutronization burst phase can confirm or defy the mixing of $\nu_e$ and $\nu_s$ for a very large range of parameter space,
if such mixing does exist, it requires other independent method to better pin down the exact range of the parameters.


\section{Conclusions}\label{sec:conclusion}

The mixings of eV-scale sterile neutrinos with active neutrinos may explain anomalies found in experimental results, such as LSND, MiniBooNE, and the deficit of the reactor neutrino fluxes.
Meanwhile, these mixings can also affect the SN phenomenon and the associated neutrino signals. Although the flavor conversion between active and sterile neutrinos in SNe have been studied extensively in the literature, an assessment of how \emph{quantitatively} future detection of SN neutrinos by different detectors can address the issue of active-sterile mixing was still lacking.
In this work, we have computed the expected SN neutrino events in neutrino detectors, including the Hyper-K, DUNE and JUNO, during the neutronization burst phase of SN neutrino emission for cases with and without $\nu_e$--$\nu_s$ mixing, assuming a non-zero mixing angle $\theta_{14}$.
We showed for the first time that by considering the total detected event numbers in all three detectors from a galactic SN neutrinos, it allows to place very stringent bounds on the $\nu_e$--$\nu_s$ mixing parameter space.

Specifically, for a SN occurring at $\lesssim 10$~kpc away from the Earth, regions where $\sin^2 2\theta_{14} \gtrsim 10^{-5}\times \sqrt{{\rm eV}^2/\Delta m^2_{41}}$ can be robustly excluded at $\geq 99\%$~C.L., for either the normal or inverted mass ordering (see Fig.~\ref{Fig:Constraint_CL}).
The key to obtain strong bounds in either mass ordering is to detect imminent SN events in multiple detectors, such as ArCC events in DUNE (specific to $\nu_e$), $e$ES events in Hyper-K (detect $\nu_e$, $\nu_x$, $\bar{\nu}_e$, $e$ES) and $p$ES events in JUNO (detect $\nu_e$, $\bar{\nu}_e$, $\bar{\nu}_x$, and $\bar{\nu}_x$). Particularly for the case of normal mass ordering, the NC detection in Hyper-K and JUNO play the most crucial roles. 
The projected constraint will not only be complementary to bounds obtained from terrestrial experiments mostly at large $\sin^2 2\theta \gtrsim 10^{-3}$, but also can exceed the existing bound derived using the Planck data by roughly two orders of magnitude (see Fig.~\ref{Fig:Constraint_compare}).

We have also shown that the projected bounds are insensitive to various uncertainties such as the neutrino fluxes predicted by SN simulations and the systematics of the detectors (see Sec.~\ref{sec:fluuncer} and \ref{sec:systemerror}).
However, if the $\nu_e$--$\nu_s$ mixing does exists, using the total neutrino events alone cannot provide precise information about the mixing parameters (see Sec.~\ref{sec:discovery}).
Nevertheless, this aspect may be addressed by considering the spectral shape or the time profile of the detected neutrino events during the SN neutronization burst (e.g., the delayed-peak due to the kinematic effect as a result of sterile neutrinos having heavier masses discussed in Ref.~\cite{Esmaili:2014gya}).
Such exploration is, however, beyond the scope of this work.
%
Further detailed analysis adopting improved statistical methods and detector simulations can be pursued to solidify the conclusions of this work.

Throughout this work, we have opted to solely focus on the neutronization burst phase and left out the neutrino signals in later phases of accretion and PNS cooling.
Although one expects to record orders of magnitude more events during these two phases, the larger uncertainties in the predicted neutrino fluxes and the progenitor dependence, as well as the yet-unclear picture of collective neutrino oscillations among active flavors due to the non-linear nature of the neutrino--neutrino interaction may prevent us from obtaining robust conclusion in probing the mixing between sterile and active neutrinos.
Moreover, the interplay of other mixing angles, $\theta_{24}$, $\theta_{34}$, and the CP phases can also affect the predicted events.
Future work are needed to elucidate these issues.

Last but not the least, it will also be of interest to examine additional scenarios beyond the simple $3+1$ scheme which might help resolve the tension between the appearance and disappearance data sets or reconcile light sterile neutrinos with cosmology. This includes, for example, the  $3+2$ $\nu$-mixing scheme, the neutrino invisible decays, or secret interaction among sterile neutrinos. 
%
%
This work and the future extension thus highlight the exquisite capability of future SN neutrino detection on probing the fundamental properties of neutrinos.

\acknowledgments
We thank Hui-Ling Li, Yu-Feng Li, and Gang Guo for helpful discussions and two anonymous JUNO internal reviewers for their valuable comments to this work.
This work was supported in part by Guangdong Basic and Applied Basic Research Foundation under Grant No. 2019A1515012216. The work was also supported in part by NSFC grant 11505301. JT acknowledge the support from the CAS Center for Excellece in Particle Physics (CCEPP). TCW was partially supported by Postdoctoral recruitment program in Guangdong province. 
MRW acknowledges supports from the Academia Sinica under Grant No. AS-CDA-109-M11, the Ministry of Science and Technology, Taiwan under Grant No. 108-2112-M-001-010, and the Physics Division, National Center of Theoretical Science of Taiwan.

\begin{appendix}

\section{flavor conversions for the inverted ordering}
\label{sec:flacon_IO}

For the IO, the normalized neutrino fluxes, $F_{\nu_\alpha}$, taking into account the flavor conversions in $3+1$ scheme are related to the fluxes without oscillations, $F^0_{\nu_\alpha}$, by
\begin{subequations}\label{eq:Fe_inverted}
\begin{align}
F_{\nu_e}= &
\left\{|U_{e2}|^2p_{14}+|U_{e4}|^2(1-p_{14})\right\}F_{\nu_e}^0
+\left(|U_{e1}|^2+|U_{e3}|^2\right)F_{\nu_x}^0, \\
F_{\nu_x}= & (|U_{\mu 2}|^2+|U_{\tau 2}|^2)p_{14}F_{\nu_e}^0+\left(|U_{\mu 1}|^2+|U_{\tau 1}|^2+|U_{\mu 3}|^2+|U_{\tau 3}|^2\right)F_{\nu_x}^0, \\
F_{\bar{\nu}_e}= & 
|U_{e3}|^2F^0_{\bar{\nu}_e}+\left(|U_{e1}|^2+|U_{e2}|^2\right)F^0_{\bar{\nu}_x}, \\
F_{\bar{\nu}_x}= & (|U_{\mu 3}|^2+|U_{\tau 3}|^2)F^0_{\bar{\nu}_e}+\left(|U_{\mu 1}|^2+|U_{\tau 1}|^2+|U_{\mu 2}|^2+|U_{\tau 2}|^2\right)F^0_{\bar{\nu}_x}.
\end{align}
\end{subequations}

For the 3-$\nu$ mixing scenario, the above equations reduce to
\begin{subequations}\label{eq:Fe_inverted_3nu}
\begin{align}
F_{\nu_e}= &
|U_{e2}|^2 F^0_{\nu_e}+ (1-|U_{e2}|^2) F^0_{\nu_x}, \\
F_{\nu_x}= & F_{\nu_x}=(1-|U_{e2}|^2) F^0_{\nu_e}+ (1+|U_{e2}|^2) F^0_{\nu_x}, \\
F_{\bar{\nu}_e}= & 
|U_{e3}|^2F^0_{\bar{\nu}_e}+(1-|U_{e3}|^2)F_{\bar{\nu}_x}^0, \\
F_{\bar{\nu}_x}= & F^0_{\bar{\nu}_e}+(1+|U_{e3}|^2)F_{\bar{\nu}_x}^0.
\end{align}
\end{subequations}


\end{appendix}

%
%
   \bibliographystyle{JHEP} 
   \bibliography{paper.bib} 
\end{document}